\documentclass[final,leqno,onefignum,onetabnum]{siamltex1213}

\usepackage{amsmath}
\usepackage{amssymb}
\usepackage{amsfonts}
\usepackage{graphicx}
\usepackage{dcolumn}
\usepackage{bm}
\usepackage{mathabx}
\usepackage{color}

\DeclareGraphicsExtensions{.pdf,.png,.jpg,.eps}

\def\Det{\textrm{Det}}

\def\p{^{\prime}}
\def\bcero{\bn{0}}
\newcommand{\bn}[1]{\mbox{\boldmath $#1$}}

\def\bB{\bn{B}}
\def\bP{\bn{P}}
\def\bY{\bn{Y}}

\def\bW{\bn{W}}
\def\bG{\bn{G}}
\def\bI{\bn{I}}
\def\bA{\bn{A}}

\def\bK{\bn{K}}

\def\bQ{\bn{Q}}
\def\bH{\bn{H}}
\def\bT{\bn{T}}
\def\bE{\bn{E}}
\def\bR{\bn{R}}
\def\bS{\bn{S}}
\def\bX{\bn{X}}
\def\bZ{\bn{Z}}
\def\bF{\bn{F}}
\def\bU{\bn{U}}
\def\bu{\bn{u}}
\def\bk{\bn{k}}
\def\bTheta{\bn{\Theta}}

\def\bUpsilon{\bn{\Upsilon}}

\def\bEta{\bn{\eta}}
\def\bmu{\bn{\mu}}
\def\brho{\bn{\rho}}

\title{Relations between transfer matrices and numerical stability analysis
to avoid the $\Omega d$ problem \thanks{This work was supported by
CONACyT (R. P.-S.), CONACyT grant 208108 (R. P.-A.) and the
Spanish Ministerio de Ciencia y Econom\'{\i}a through grant
MAT2012-38045-C04-04 (V. R. V.)}}

\author{R. P\'{e}rez-\'Alvarez
\thanks{Universidad Aut\'{o}noma del Estado de Morelos, Ave. Universidad 1001, CP 62209, Cuernavaca, Morelos,
M\'{e}xico} \and R. Pernas-Salom\'on
\thanks{Universidad Aut\'{o}noma del Estado de Morelos, Ave. Universidad 1001, CP 62209, Cuernavaca, Morelos,
M\'{e}xico} \and V. R. Velasco
\thanks{Instituto de Ciencia de Materiales de Madrid (ICMM), Consejo Superior de Investigaciones Científicas (CSIC),
c/ sor Juana In\'es de la Cruz 3, 28049 Madrid, Spain}}

\begin{document}
\maketitle
\slugger{mms}{xxxx}{xx}{x}{x--x}

\begin{abstract}
The transfer matrix method is usually employed to study problems
described by $N$ equations of matrix Sturm-Liouville (MSL) kind.
In some cases a numerical degradation (the so called $\Omega d$
problem) appears thus impairing the performance of the method. We
present here a procedure that can overcome this problem in the
case of multilayer systems having piecewise constant coefficients.
This is performed by studying the relations between the associated
transfer matrix ($\bT$) and other transfer matrix variants. In
this way it was possible to obtain the matrices which can overcome
the $\Omega d$ problem in the general case and then in problems
which are particular cases of the general one. In this framework
different strategies are put forward to solve different boundary
condition problems by means of these numerically stable matrices.
Numerical and analytic examples are presented to show that these
stable variants are more adequate than other matrix methods to
overcome the $\Omega d$ problem. Due to the ubiquity of the MSL
system, these results can be applied to the study of many
elementary excitations in multilayer structures.
\end{abstract}

\begin{keywords}
Transfer matrix, matrix Sturm-Liouville problem, numerical
stability, quadratic eigenvalues, $\Omega d$ problem
\end{keywords}

\begin{AMS}
34L16
\end{AMS}

\pagestyle{myheadings} \thispagestyle{plain} \markboth{R.
P\'{E}REZ-\'ALVAREZ, R. PERNAS-SALOM\'ON AND V. R.
VELASCO}{TRANSFER MATRIX NUMERICAL STABILITY AND THE $\Omega d$
PROBLEM}

\date{\today }


\maketitle

\section{Introduction}

The study of elementary excitations in multilayer systems
(heterostructures) continues to be a very active field of research
due to the multiple applications of these systems for the design
of devices with composite materials. In recent years
magneto-electro-elastic materials \cite{2012-YuJiangong-AMM} and
piezoelectric multilayer structures \cite{2009-Guo-SchSG,
2010-HCalas-JAP}, among other systems, have been the object of
many studies. The associated transfer matrix method
\cite{LibroUJI} $\bT$ is one of the theoretical techniques most
employed in the study of these systems. From the formal point of
view this method is very adequate for the study of problems
related with multilayer systems. It reflects in a very simple way
the linearity of the problem, based on the fact that any solution
can be expressed as a linear combination of a chosen basis of the
corresponding functional space \cite{LibroUJI}. On the other hand,
for several practical applications to different problems, the
method is hampered by numerical instabilities, the most common one
being called the $\Omega d$ problem \cite{LibroUJI}. The name associated to this
numerical instability derives from the elastic waves studies where
this instability is present at high frequencies $\Omega$ and/or
big thicknesses $(d)$ of the layers.

A description of this problem was given in \cite{1965-Dunkin-BSSA}
when studying wave propagation in layered elastic media at high
frequencies. In this work the origin of the problem was assigned
to the large frequency-thickness ($fd$) products. It was found
that the modal calculations presented numerical difficulties and a
matrix formulation, the $\delta$-matrix method, was proposed to
deal with them. Another approach closely related with the
scattering matrix method is the reflectivity matrix method
\cite{1992-Rokhlin-JAcSA}. Nevertheless the high
frequency-thickness product instability has been a persistent
feature in the study of wave propagation in layered media as it
can be seen in a later review \cite{1995-Lowe-IEEETrans}. The name
$\Omega d$ problem was coined in \cite{LibroUJI}. There it was
noticed that the expressions producing this instability are of the
form $\Omega d$ where $\Omega$ stands for a function of the
frequency or the equivalent magnitude for other elementary
excitations.

Different techniques have been developed to deal with this
problem. Some of them, as the global transfer matrix
\cite{LibroUJI, 2010-HCalas-JAP}, involve matrices with dimensions
increasing with the number of layers forming the system. It is
then clear that for systems including many layers the method will
require big amounts of computer memory and time.

Other approaches employ transfer matrices with dimensions
independent of the number of layers. Among them we can find the
Stiffness matrix method \cite{2002-SIRokhlin-JASA,
2004-LWang-IEEE51, 2005-ELTan-ASA, 2007-ELTan-IEEE} $(\bE)$, the
Scattering matrix method \cite{2007-ELTan-IEEE, 2003-ELTan-Ultras,
2011-AAShcherbakov-OE} $(\bS)$ and the method of the hybrid
compliance-stiffness (or simply hybrid matrix)
\cite{2007-ELTan-IEEE, 2006-ELTan-ASA, 2010-ELTan-Ultras} $(\bH)$.
These methods have been mainly used in studies of elastic waves
propagation in anisotropic systems and acoustic waves in piezoelectric systems.

All these studies have been performed in a separate way. There is
no clear picture of the usefulness and limitations of the
different approaches. Our aim is to give an unified view of the
problem and present the most adequate transfer matrix variant to
solve different problems. To this end we shall consider a general
system of $N$ differential equations of the matrix Sturm-Liouville
(MSL) kind \cite{LibroUJI}. In this general framework the study of
the expressions relating each different matrix with $\bT$ will
allow to understand how the different matrices elude the $\bT$
numerical instabilities. To extend the use of these transfer
matrix variants $\bH$, $\bE$ and $\bS$ to a wider range of
physical problems involving multilayer systems we shall study the
numerical stability of each matrix variant. In addition we shall
present different strategies to be used in the case of common
boundary value problems as superlattices or finite sandwiches in
terms of $\bH$, $\bE$ or $\bS$.

We must stress that our procedure can overcome this problem in the
case of multilayer systems having piecewise constant coefficients.
This is an important problem and covers many cases of practical
interest.

Among the big amount of work done on this problem using other
methods we can mention those based on the sextic formalism for the
linear elasticity \cite{2003-Shuvalov-PRSL}. In this scheme the
matricant matrix was introduced together with the impedance matrix
and the two point impedance matrix \cite{2010-Norris-QJMAP}. This
approach allows to deal with systems having inhomogeneous
coefficients. Stable methods to compute the matricant and the
impedance matrix with special integration schemes
\cite{2013-Norris-JSV} and an alternative method based on the
resolvent of a propagator have been presented recently
\cite{2013-Korotyaeva-APL, 2014-Korotyaeva-JCA}. In these works
the chain rule for the resolvent, together with a differential
equation of Riccati kind for the obtention of the resolvent in
continuous inhomogeneous media are presented. The resolvent is
well adapted to get the spectrum and fields in these systems.

The general character of our approach allows the extension of the
transfer matrix variants use to problems whose systems of
equations are particular cases of the MSL. We shall illustrate,
for example, the hybrid matrix numerical stability by numerical
studies of the shear horizontal surface waves in piezoelectric
multilayer systems.

In Section \ref{SLM} we present the master equation of the matrix
Sturm-Liouville system of equations. In Section \ref{QEP} we
introduce the quadratic eigenvalues problem leading to the
linearly independent solutions of the system together with their
eigenvalues for a homogeneous medium. We define in Section
\ref{ATM} the associated transfer matrix $\bT$ and introduce the
form employed in the analysis of the numerical stability of the
variants $\bH$, $\bE$ and $\bS$. Section \ref{Problema-Omega-d}
introduces the $\Omega d$ problem together with the analysis of
the $\bT$ characteristics that can be the source of this numerical
instability. Afterwards, the numerical stability of the hybrid
matrix, Section \ref{estabilidad-H}, and the stiffness matrix,
Section \ref{estabilidad-E}, are studied through their respective
relations with $\bT$ in an homogeneous domain. The analysis for
the scattering matrix is presented in (Section
\ref{estabilidad-S}). The composition rules for the different
matrices considered here are analyzed in Section \ref{ReglasdeC}.
Section \ref{aplicaciones} presents the strategies to solve
several boundary problems in terms of $\bH$, $\bE$ or $\bS$. A
numerical example demonstrating the numerical stability of the
hybrid matrix is also presented together with an analytic study of
the well known Kronig-Penney model. Conclusions are presented in
Section \ref{Conclusiones}.

\section{\label{SLM}Matrix Sturm-Liouville system of equations (MLS)}

A matrix Sturm-Liouville problem emerges naturally in a wide range
of physical and technological problems (see, for example Refs.
\cite{LibroUJI, CubaLibro, 2001-FTisseur-SIAM}, and citations
therein). In this wide range of problems there are many belonging
to the elasticity theory (see for example
\cite{2009-GBonnet-IJSS}), electromagnetism
\cite{2007-XiangyongLi-IJSS} and several other areas of classical
physics. Some of these problems can be quite complicated as the
magneto-electro-elastic waves \cite{2007-JiangyiChen-IJSS}. A
matrix Sturm-Liouville problem appears also in Quantum Mechanics
and Solid State Physics. Particularly the Envelope Function
Approximation (EFA) \cite{1986-GBastard-IEEEQE, Bastard89}
generates a massive class of systems of equations that follow the
Sturm-Liouville equation in matrix form. Initially many of these
systems of equations are three-dimensional, but in layered
systems, as outlined in Figure \ref{Figura01}, the normal modes
can be chosen as exponential of $i\vec{\kappa}\cdot\vec{\rho}$
multiplied by some function of the variable $z$, the coordinate
perpendicular to the interfaces. We denote by $\vec{\rho} = x
\vec{e}_x + y \vec{e}_y$ the position vector in the plane of the
interfaces and  by $\vec{\kappa} = \kappa_x \vec{e}_x + \kappa_y
\vec{e}_y$ the corresponding wavevector. In this way the equations
of motion take the Sturm-Liouville form, namely:

\begin{eqnarray}
\label{Eqmaestra} \frac{d}{dz} \left[ \bB(z) \cdot \frac{d\mathbf{F}(z)}{dz} + \bP(z) \cdot \mathbf{F}(z)\right] + \bY(z)
\cdot \frac{d\mathbf{F}(z)}{dz} + \bW(z) \cdot \mathbf{F}(z) & = & \bcero \;.
\end{eqnarray}

This defines the matrix differential operator $\mathbf{L}(z)$.The
unknown $\mathbf{F}(z)$ is the field under study: electronic
wavefunctions, or envelope functions, if we deal with electronic
states, vibration amplitude for elastic waves, or components of
the electric field in some electrodynamic situations. In the case
of the Full Phenomenological Model (FPM) for polar optical modes
in heterostructures \cite{CubaLibro} the unknown field has several
components: three mechanical amplitudes and a component which is
interpreted as a coupled electrostatic potential \cite{CubaLibro}.
The coefficients $\bB(z)$, $\bP(z)$, $\bY(z)$, and $\bW(z)$ are
square matrices of order $N$, being $N$ the number of coupled
second order differential equations forming the system
(\ref{Eqmaestra}). These coefficients characterize the physical
properties of the materials forming the multilayer system:
dielectric constants, elastic coefficients, etc. As the multilayer
structures studied here involve different materials these
coefficients will be different for the different materials. The
dot $\cdot$ means standard matrix product.

As the linear differential form is defined from (\ref{Eqmaestra})

\begin{eqnarray}
  \mathbf{A}(z) &=& \bB(z) \cdot \frac{d\mathbf{F}(z)}{dz} + \bP(z) \cdot \mathbf{F}(z),
\end{eqnarray}

\noindent then the first integration from $z-\epsilon$ to
$z+\epsilon$ shows that $\mathbf{A}(z)$ is continuous for every
$z$ along the multilayer structure. The continuity of
$\mathbf{F}(z)$ and $\mathbf{A}(z)$ along the structure allows to
obtain the composition rule for the transfer matrices defined from
these magnitudes.

\subsection{\label{QEP} LI solutions. Quadratic Eigenvalues Problem}

In the case of an homogeneous medium the differential equations system (\ref{Eqmaestra}) takes the following form

\begin{eqnarray}
\label{Ec-maestra-homog} \bB\cdot \mathbf{F}^{\prime\prime}(z) +(\bP+\bY)\cdot\; \mathbf{F}\p(z)
+\bW\cdot\mathbf{F}(z)&=& \bcero\;.
\end{eqnarray}
In this simple case the linearly independent (LI) solutions of the
differential equations system (\ref{Eqmaestra}) can be expressed
by means of exponentials \cite{Hurewicz, Bibikov}

\begin{eqnarray}
\label{Soluciones-LI-general} \mathbf{F}(z) &=& \mathbf{F}_{0} \,
e^{ik\, z}\;.
\end{eqnarray}

The eigenvalues $k$ are obtained from the zeros of the secular matrix determinant:

\begin{eqnarray}
\label{2003.08.28c} \bTheta(k) &=& -k^2 \bB + i k (\bP+\bY) + \bW \;.
\end{eqnarray}

Now we are dealing with a quadratic eigenvalues problem (QEP)
\cite{2001-FTisseur-SIAM}. If matrix $\bB$ is \textit{regular}
$(\Det[\bB]\neq0)$ we have a set of eigenvalues $K=\{k_j,
j=1,2,\cdots,2N\}$ and the corresponding eigenfunctions
$\mathbf{F}_j(z) = \mathbf{F}_{j0} \, \exp[ik_jz]$. The amplitudes
$\mathbf{F}_{j0}$ multiplied by a constant are obtained from the
homogeneous linear equations system:

\begin{eqnarray}
\label{2003.08.28d} \bTheta(k_j)\cdot\mathbf{F}_{j0} &=& \bcero \;.
\end{eqnarray}

The multiplicative constant is usually obtained by a normalization condition.

In the following, we shall always assume $\bB^\dag=\bB$,
$\bW^\dag=\bW$ and $\bY=-\bP^\dag$, in order to ensure formal
hermiticity of the operator $\mathbf{L}(z)$, see Ref.
\cite{LibroUJI}.  In this case the eigenvalues of the QEP satisfy
the general property of being real or appearing in pairs: $k_j$
and its complex conjugate $k_j^*$.

\subsection{\label{ATM} Associated Transfer Matrix for the MSL equations system}

We shall define the associated transfer matrix
$\bT(\alpha:z,z_0)$, which transfers the amplitudes $\mathbf{F}$
and the linear differential form $ \mathbf{A}$ in a domain
$\alpha$, as in \cite{LibroUJI}:

\begin{eqnarray}
\label{matriz-T}
  \begin{array}{|c|}
     \mathbf{F}(\alpha:z) \\
      \mathbf{A} (\alpha:z) \\
    \end{array}
   &=& \bT (\alpha:z,z_0)\;\cdot\;
                             \begin{array}{|c|}
                                \mathbf{F}(\alpha:z_0) \\
                                \mathbf{A} (\alpha:z_0)\\
                             \end{array}.
\end{eqnarray}

From now on we shall suppress the zonal argument $\alpha$.
Following the algebraic and analytic methods to calculate the
matrix $\bT(z,z_0)$, given in \cite{LibroUJI}, we shall have:

\begin{eqnarray}\label{T-Q}
  \bT(z,z_0) &=& {\bQ}(z)\cdot {\bQ}(z_0)^{-1},
\end{eqnarray}

\noindent where the auxiliary matrix $\bQ(z)$ is formed by a basis
of eigenfunctions $\mathbf{F}_j(z)$ and of the linear differential
forms $\mathbf{A}_j (z)=\bB(z) \cdot
\displaystyle{\frac{d\mathbf{F}_j(z)}{dz}} + \bP(z) \cdot
\mathbf{F}_j(z)$:

\begin{eqnarray}
\label{Q-ConsT}
 \bQ(z) &=& \left|\begin{array}{cccc}
                    \mathbf{F}_1(z) &  \mathbf{F}_2(z) & \ldots &  \mathbf{F}_{2N}(z) \\
                    \mathbf{A}_1 (z) & \mathbf{A}_2 (z) & \ldots & \mathbf{A}_{2N} (z)
                  \end{array}\right|.
\end{eqnarray}

For an homogeneous domain $\alpha$, with constant $\bB$, $\bP$ and
$\bW$, we can choose the eigenfunctions
$\mathbf{F}_j(z)=\mathbf{F}_{j0}\;e^{ik\, z}$, and after some
manipulations on (\ref{Q-ConsT}), we can separate the factors
$\mathbf{F}_{j0}$ from the exponentials $e^{ik\, z}$ in the form:

\begin{eqnarray}
\label{Q-ConstT4}
  \bQ(z) &=& \left[
                   \begin{array}{cc}
                      \mathbf{F}_{0_N} &\; \mathbf{F}_{0_{2N}}\\
                      \mathbf{A}_{0_N} &\;\mathbf{A}_{0_{2N}} \\
                   \end{array}
                 \right]\cdot \left[\begin{array}{cc}
                               \mathbf{ \Pi}_{k_N}(z-z_0) & \bcero \\
                                \bcero & \mathbf{\Pi}_{k_{2N}}(z-z_0)
                              \end{array}\right] \cdot \left [ \begin{array}{cc}
                                \mathbf{\Pi}_{k_N}(z_0) & \bcero \\
                                \bcero  & \mathbf{\Pi}_{k_{2N}}(z_0)
                              \end{array}\right].\nonumber
\end{eqnarray}

The submatrices $\mathbf{\Pi}_{k_N}(d)$ and
$\mathbf{\Pi}_{k_{2N}}(d)$ are diagonal an the $j$\-th element is
the exponential $e^{ik_j\;d}$. $\mathbf{F}_{0_N}$,
$\mathbf{A}_{0_N}$, $\mathbf{F}_{0_{2N}}$ and
$\mathbf{A}_{0_{2N}}$ are square matrices of order $N$ whose
elements are obtained in terms of the constant $\mathbf{F}_{j0}$
and the corresponding $\mathbf{A}_{j0}$. In our notation the
subindex $\{N\}$ denotes that $j=1,2,\cdots,N$ and the subindex
$\{2N\}$ means that $j=N+1,N+2,\cdots,2N$.

By substituting $\bQ(z)$ in (\ref{T-Q}) and considering $d$=$z-z_0$ we have:

\begin{eqnarray}
\label{T-d}
   \bT(d) &=& \left[
                   \begin{array}{cc}
                       \mathbf{F}_{0_N} &\;  \mathbf{F}_{0_{2N}}\\
                      \mathbf{A}_{0_N} &\;\mathbf{A}_{0_{2N}} \\
                   \end{array}
                 \right]\cdot \left[ \begin{array}{cc}
                                \mathbf{\Pi}_{k_N}(d) & \bcero \\
                                \bcero  & \mathbf{\Pi}_{k_{2N}}(d)
                              \end{array}\right] \cdot \left[
                   \begin{array}{cc}
                       \mathbf{F}_{0_N} &\;  \mathbf{F}_{0_{2N}}\\
                      \mathbf{A}_{0_N} &\;\mathbf{A}_{0_{2N}} \\
                   \end{array}
                 \right]^{-1}.
\end{eqnarray}

The matrix $\bT(d)$ appearing in (\ref{T-d}) can be interpreted as
the associated transfer matrix (ATM) relating the vector
$\left[\mathbf{F}(z)\;\; \mathbf{A}(z)\right]^T$ in the boundaries
of an homogeneous domain with thickness $d$.

\begin{figure}
\begin{center}
\begin{picture}(300,250)(-30,-30)
\put(-100,0){\vector(1,0){400}}
\put(-60,30){\line(0,1){70}}
\put(-40,30){\line(0,1){70}}
\put(-10,30){\line(0,1){70}}
\put(10,30){\line(0,1){70}}
\put(60,30){\line(0,1){70}}
\put(90,30){\line(0,1){70}}
\put(122,30){\line(0,1){70}}
\put(170,30){\line(0,1){70}}
\put(220,30){\line(0,1){70}}
\put(252,30){\line(0,1){70}}
\put(274,30){\line(0,1){70}}
\put(300,-10){$z$} \put(-80,60){${\rm L}$} \put(-64,-10){$z_\ell$} \put(-62,10){$\vdots$} \put(-50, 60){$1$}
\put(-44,-10){$z_1$} \put(-42,10){$\vdots$} \put(-30,60){$2$} \put(-14,-10){$z_2$} \put(-12,10){$\vdots$}
\put(0,60){$3$} \put(06,-10){$z_3$} \put(08,10){$\vdots$} \put(22,60){$\cdots$} \put(22,-10){$\cdots$}
\put(56,-10){$z_{m-1}$} \put(58,10){$\vdots$} \put(65,60){$m$} \put(86,-10){$z_{m}$} \put(88,10){$\vdots$}
\put(91,60){$m+1$} \put(116,-10){$z_{m+1}$} \put(120,10){$\vdots$} \put(145,-10){$\cdots$} \put(132,60){$\cdots$}
\put(164,-10){$z_{\mu-3}$} \put(168,10){$\vdots$}
\put(180,60){$\mu-2$} \put(214,-10){$z_{\mu-2}$} \put(218,10){$\vdots$} \put(223,60){$\mu-1$}
\put(245,-10){$z_{\mu-1}$} \put(251,10){$\vdots$} \put(260, 60){$\mu$} \put(273,-10){$z_r$} \put(273,10){$\vdots$}
\put(280, 60){${\rm R}$}
\end{picture}
\caption{\label{Figura01} General scheme of the system under
study. The system consists of $\mu$ layers sandwiched by two
semi-infinite external domains: ${\rm L}$ (left) and ${\rm R}$
(right). According to our convention, the layer $m$ is bounded
between interfaces $(m-1)$ and $m$ with coordinates $z_{m-1}$ and
$z_m$ respectively.}
\end{center}
\end{figure}
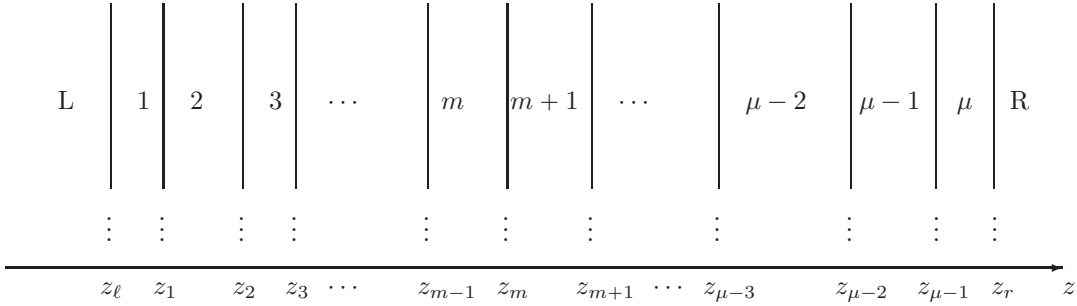

As the linear form $\mathbf{A}(z)$ is continuous along the
interface separating two adjacent domains, the ATM has the chain
property. Then for an ensemble of $\mu$ layers sketched in Figure
\ref{Figura01}, the system ATM is obtained from the following
matrix product:

\begin{eqnarray}
  \bT (z_r, z_\ell) &=& \bT (z_r-z_{\mu-1})\ldots \bT (z_2-z_1)\cdot\bT (z_1-z_\ell),
\end{eqnarray}

\noindent where $z_\ell,z_1,z_2,\ldots,z_r $ are coordinates of
the interfaces matching the different domains of the multilayer
structure.

We shall start now the study of the $\bT$ characteristics which
can be the source of the $\Omega d$ problem in the numerical
calculations. With this knowledge we shall study later the
numerical stability of the $\bH$, $\bE$ and $\bS$ matrices, by
means of their relations with $\bT$.

\subsection{\label{Problema-Omega-d}$\Omega d$ problem}

From (\ref{T-d}) we can obtain expressions for the analysis of the
numerical instability of the $\bT$ matrix elements for any $N$.
For real eigenvalues (allowed regions) we have:

\begin{eqnarray}
   \bT_{ls} &=& \sum_{j=1}^{2N} A_{lsj}\left[\cos (k_j\;d)+i \sin(k_j\;d)\right],
\end{eqnarray}

\noindent whereas for complex eigenvalues (forbidden regions) we
shall have combinations of decreasing and increasing exponentials:

\begin{eqnarray}
\label{T-ls-complejo}
  \bT_{ls} &=& \sum_{j=1}^{N} C_{lsj}\;\tau_j\; e^{|\Im(k_j)|\;d}\left(1\pm D_{lsj} e^{-2|\Im(k_j)|\;d}\right).
\end{eqnarray}

The coefficients $A_{lsj}$, $C_{lsj}$ and $D_{lsj}$ are expressed
in terms of the elements of $\mathbf{F}_{0_{N}}$,
$\mathbf{A}_{0_{2N}}$, $\mathbf{F}_{0_{2N}}$ and
$\mathbf{A}_{0_{2N}}$. In (\ref{T-ls-complejo}) we have separated
the $k_j$ eigenvalue real $\Re(k_j)$ and imaginary $\Im(k_j)$
parts. The real part is included in the $\tau_j=e^{i\Re(k_j)d}$
factor having a bounded value.

For real eigenvalues the $\bT$ elements are represented by means
of trigonometric functions which are bounded by $\pm1$. In this
case the $\Omega d$ problem does not appear when the product
$(k_j\;d)$ increases. On the other hand, for complex eigenvalues
the mixing of terms with increasing and decreasing exponential
values present in (\ref{T-ls-complejo}) may give rise to this
numerical instability.

For increasing $k_j\;d$ leading to $D_{lsj}
e^{-2|\Im(k_j)|\;d}\approx u$ (\textit{unit roundoff}) the
$\left(1\pm D_{lsj} e^{-2|\Im(k_j)|\;d}\right)$ operation is
rounded to $1.0$ by the computer. Thus the result $(1.0)$ will
have a round-off error. The number $u$ (\textit{unit roundoff}) is
the machine precision, that is, the value to be added to $1.0$ to
produce a result different from $1.0$. This number can be
calculated as $u=\frac{1}{2}\beta^{1-t}$ \cite{Higham-2002-ASN},
where $\beta$ is the base of the floating point number system and
$t$ its precision (can be understood as the number of digits used
to give a value). The roundoff error is defined as the difference
between the calculated approximation of a number and its exact
mathematical value. When the roundoff result is 1.0 the absolute
value of this error $E_{abs}$ is bounded $E_{abs}\leq u$
\cite{Higham-2002-ASN}. In the double precision decimal system
($\beta=10$, $t=16$) we have $u=5 \times 10^{-16}$.

If we assume that the calculation of a term $C_{lsj}\;\tau_j\;
e^{|\Im(k_j)|\;d}\left(1\pm D_{lsj} e^{-2|\Im(k_j)|\;d}\right)$ of
(\ref{T-ls-complejo}) is performed with roundoff, then the result
will be affected by a roundoff error with absolute value $E_r$
given by:

\begin{eqnarray}\label{E-absoluto}
 E_r &\leq& C_{lsj}\;\tau_j\; e^{|\Im(k_j)|\;d}\;u.
\end{eqnarray}

Depending on the numerical problem under study the right-hand side
in (\ref{E-absoluto}) can have a high value and also a big $E_r$
error. The roundoff error can be accumulated when the final result
to be obtained (e.g., eigenvalues or parameters of a given
problem) is preceded by a sequence of calculations prone to
roundoff errors. In these cases the error can dominate the
calculations thus giving a very inaccurate final result. When this
happens we are in the presence of the numerical instability called
$\Omega d$ problem.

In practice it is quite easy to deal with problems in which the
$\bT$ determinant is constant and equal to one. This can be used
as a test of the numerical accuracy in the real calculations. When
the numerical instability is present the $\Det[\bT]$ takes values
quite different from the exact one, being in some cases several
orders of magnitude bigger or smaller than 1.0.

The expression (\ref{T-ls-complejo}) shows also clearly that the
$\bT_{ls}$ elements increase indefinitely when exponential
argument $|\Im(k_j)|\;d\rightarrow \infty$. In this case the $\bT$
matrix overflows and cannot be calculated numerically. Thus it is
clear that in the $\bT$ numerical applications we can find two
kinds of numerical instability: the $\Omega d$ problem and the
matrix overflow.

\section{\label{H-E}Hybrid matrix and Stiffness matrix of the MSL system}

We can define new matrices in the domain $\alpha$, where $\bT$ was
defined, by changing the arrangement of the $\mathbf{F}(z)$,
$\mathbf{A}(z)$, $\mathbf{F}(z_0)$ and $ \mathbf{A}(z_0)$ vectors
in (\ref{matriz-T}). Some examples are the Hybrid
Compliance-Stiffness matrix $(\bH)$ and the Stiffness matrix
$(\bE)$:

\begin{eqnarray}
\label{matriz-H-1}
  \begin{array}{|c|}
    \mathbf{F}(\alpha:z_0) \\
     \mathbf{A} (\alpha:z)
   \end{array}
   &=& \bH(\alpha:z;z_0)\;\cdot \begin{array}{|c|}
                           \mathbf{A} (\alpha:z_0) \\
                           \mathbf{F}(\alpha: z)
                         \end{array}\;.
\end{eqnarray}

\begin{eqnarray}
\label{matriz-E}
  \begin{array}{|c|}
      \mathbf{A} (\alpha:z_0) \\
      \mathbf{A} (\alpha:z)
   \end{array}
   &=& \bE(\alpha:z;z_0)\;\cdot \begin{array}{|c|}
                            \mathbf{F}(\alpha:z_0) \\
                            \mathbf{F}(\alpha:z)
                         \end{array}\;.
\end{eqnarray}

The Hybrid Compliance-Stiffnes matrix was employed in Ref.
\cite{2006-ELTan-ASA} as a stable variant to study the propagation
of an acoustic wave in an anisotropic multilayer system. The
acoustic wave equations of motion are a particular case of the
system (\ref{Eqmaestra}) including the displacement vector as
$\mathbf{F}(z)$ and the force vector normal to the interfaces as
$\mathbf{A}$(z).

Following this procedure we can define up to 24 interrelated
matrices related among them. In fact we obtain 12 different
matrices and their respective inverses. Among them we find
$\bT^{-1}$, $\bH^{-1}$ and $\bE^{-1}$. The matrix $\bE^{-1}$is
known as Compliance matrix, see Refs. \cite{2006-ELTan-ASA,
2002-SIRokhlin-JASA}. By taking as reference the expressions
defining $\bT$, $\bH$, $\bE$, $\bT^{-1}$, $\bH^{-1}$ and
$\bE^{-1}$ is possible to obtain from them other three different
matrices which will exhibit a similar numerical behaviour. A first
matrix is obtained by permuting among them the positions of the
vectors in the right-hand side of the matrix taken as reference
(e.g., the $\mathbf{A} (\alpha:z_0)$ and $\mathbf{F}(\alpha: z)$
vectors in the right-hand side of (\ref{matriz-H-1}). The second
matrix is obtained by means of the former operation applied to the
vectors in the left-hand side of the matrix taken as reference.
The third one is the result of both permutations. The Appendix
\ref{App-A} shows, by means of the relations between the matrices,
that the matrices defined in this way will have a similar
behaviour from the numerical point of view.

We obtain $\bT(-d)$ by inversion of (\ref{T-d}). Thus $\bT^{-1}$
will have the same numerical behaviour than $\bT$. When inverting
the expressions for $\bH(d)$ and $\bE(d)$ the result is the
permutation of the $\mathbf{F}_{0_N}$ submatrix with the
$\mathbf{A}_{0_N}$ one and of the $\mathbf{F}_{0_{2N}}$ submatrix
with the $\mathbf{A}_{0_{2N}}$ one. Thus $\bH^{-1}$ and $\bE^{-1}$
will have also a numerical behaviour similar to those of their
counterparts.

\subsection{\label{estabilidad-H}Analysis of the numerical instability of the Hybrid matrix of the MSL system}

The following relations can be obtained from Eqs. (\ref{matriz-T}) and (\ref{matriz-H-1}):

\begin{eqnarray}
\label{H-T}
 \bH &=& \left[\begin{array}{cc}
           -[\bT_{11}]^{-1}\cdot \bT_{12} & \;\;\;[\bT_{11}]^{-1} \\
           \bT_{22}-\bT_{21}\cdot [\bT_{11}]^{-1}\cdot \bT_{12} & \;\;\;\bT_{21} \cdot [\bT_{11}]^{-1}
         \end{array}\right],
\end{eqnarray}

On the other hand, equations (\ref{matriz-T}) and (\ref{T-Q})
exhibit an important property. Equation (\ref{T-Q}) leads to a
unique ATM independently of the chosen LI solutions base. As a
consequence the hybrid matrix obtained from the relations
(\ref{H-T}) will be independent also from the solutions base
chosen to build up $\bT$. Then, for simplicity, we consider that
the ATM expression (\ref{T-d}) was built from a base of solutions
$\mathbf{F}_j(z) = \mathbf{F}_{j0} \, e^{ik_jz}$, in such a way
that $\mathbf{\Pi}_{k_N}(d)$ contains the eigenvalues with
positive imaginary part $\Im(k_1, k_2,\ldots k_N)>0$ and
$\mathbf{\Pi}_{k_{2N}}(d)$ the eigenvalues with negative imaginary
part $\Im(k_{N+1}, k_{N+2},\ldots k_{2N})<0$. In this way the
submatrices $\mathbf{\Pi}_{k_N}(d)$ and
$\mathbf{\Pi}_{k_{2N}}(-d)$ reduce to the order $N$ nil matrix
$(\bcero_N)$ when the thickness $d\rightarrow\infty$ whereas the
elements of $\mathbf{\Pi}_{k_N}(-d)$ and
$\mathbf{\Pi}_{k_{2N}}(d)$ tend to infinity.

Appendix \ref{App-B} contains the expressions for the $N$ order
partitions: $\bT_{11}$, $\bT_{12}$, $\bT_{21}$ and $\bT_{22}$
obtained from (\ref{T-d}).With the help of (\ref{H-T}) we have:

\begin{eqnarray}\label{H11-inicial}
 \bH_{11} &=& \left[\mathbf{A}_{0_{2N}}\cdot\mathbf{F}_{0_{2N}}^{-1}
-\gamma_{21}\cdot\mathbf{\Pi}_{k_N}(-d)\cdot\mathbf{F}_{0_N}^{-1}\cdot\mathbf{F}_{0_{2N}}\cdot\mathbf{\Pi}_{k_{2N}}(d)\cdot\gamma_{12}^{-1}
\right]^{-1}+ \nonumber\\&&
\left[\mathbf{A}_{0_{N}}\cdot\mathbf{F}_{0_{N}}^{-1}-\gamma_{22}\cdot\mathbf{\Pi}_{k_{2N}}(-d)\cdot\mathbf{F}_{0_{2N}}^{-1}\cdot\mathbf{F}_{0_{N}}\cdot\mathbf{\Pi}_{k_{N}}(d)\cdot\gamma_{11}^{-1}
\right]^{-1}
\end{eqnarray}

\begin{eqnarray}\label{H12-inicial}
\bH_{12} &=&
\gamma_{12}\cdot\mathbf{\Pi}_{k_{2N}}(-d)\cdot\mathbf{F}_{0_{2N}}^{-1}\nonumber\\&&\cdot\left[\bI_N-\mathbf{F}_{0_{N}}\cdot\mathbf{\Pi}_{k_{N}}(d)\cdot\mathbf{A}_{0_{N}}^{-1}\cdot\mathbf{A}_{0_{2N}}\cdot\mathbf{\Pi}_{k_{2N}}(-d)\cdot\mathbf{F}_{0_{2N}}^{-1}
\right]^{-1}
\end{eqnarray}

\begin{eqnarray}\label{H21-inicial}
\bH_{21} &=&
\left[\mathbf{A}_{0_{N}}-\bH_{22}\cdot\mathbf{F}_{0_{N}}\right]\cdot
\mathbf{\Pi}_{k_N}(d)\cdot \gamma_{21}^{-1}+\nonumber\\&&\left[
\mathbf{A}_{0_{2N}}-\bH_{22}\cdot\mathbf{F}_{0_{2N}}\right]\cdot\mathbf{\Pi}_{k_{2N}}(d)\cdot\gamma_{22}^{-1}
\end{eqnarray}

\begin{eqnarray}\label{H22-inicial}
\bH_{22} &=& \left[\mathbf{F}_{0_{N}}\cdot\mathbf{A}_{0_{N}}^{-1}-
\mathbf{F}_{0_{2N}}\cdot\mathbf{\Pi}_{k_{2N}}(d)\cdot\mathbf{A}_{0_{2N}}^{-1}\cdot\mathbf{A}_{0_{N}}\cdot\mathbf{\Pi}_{k_{N}}(-d)\cdot\mathbf{A}_{0_{N}}^{-1}\right]^{-1}+
\nonumber\\&&\left[\mathbf{F}_{0_{2N}}\cdot
\mathbf{A}_{0_{2N}}^{-1}-\mathbf{F}_{0_{N}}\cdot\mathbf{\Pi}_{k_N}(d)\cdot\mathbf{A}_{0_{N}}^{-1}\cdot\mathbf{A}_{0_{2N}}\cdot\mathbf{\Pi}_{k_{2N}}(-d)\cdot\mathbf{A}_{0_{2N}}^{-1}
   \right]^{-1} \; .
\end{eqnarray}

The coefficients $\gamma_{11}$,  $\gamma_{12}$,  $\gamma_{21}$ and
$\gamma_{22}$ are obtained in terms of $\mathbf{F}_{0_{N}}$,
$\mathbf{F}_{0_{2N}}$, $\mathbf{A}_{0_{N}}$ and
$\mathbf{A}_{0_{2N}}$ as indicated in Appendix \ref{App-B}.

When $d$ increases indefinitely the expressions
(\ref{H11-inicial}-\ref{H21-inicial}) are reduced to:

\begin{eqnarray}
  \bH_{11}|_{d\rightarrow\infty} &=& \mathbf{F}_{0_{N}}\cdot\mathbf{A}_{0_{N}}^{-1}\cdot\left[\bI_N-\bcero_N \right]^{-1}=\mathbf{F}_{0_{N}}\cdot\mathbf{A}_{0_{N}}^{-1} \\
  \bH_{12}|_{d\rightarrow\infty} &=& \bcero_N \cdot \left[\bI_N-\bcero_N \right]^{-1}=\bcero_N \\
  \bH_{21}|_{d\rightarrow\infty}  &=& \left[\mathbf{A}_{0_{2N}}
  -\mathbf{A}_{0_{2N}}\cdot\mathbf{F}_{0_{2N}}^{-1}\cdot\left[\bI_N-\bcero_N
   \right]^{-1}\cdot\mathbf{F}_{0_{N}}\right]\cdot \bcero_N \nonumber\\&&+
   \left[\mathbf{A}_{0_{2N}}-\mathbf{A}_{0_{2N}}\left[\bI_N-\bcero_N
   \right]^{-1}\right]\cdot\mathbf{\Pi}_{k_{2N}}(d)|_{d\rightarrow\infty}\cdot\gamma_{22}^{-1}=\bcero_N
   \\
\bH_{22}|_{d\rightarrow\infty} &=&
\mathbf{A}_{0_{2N}}\cdot\mathbf{F}_{0_{2N}}^{-1}
\cdot\left[\bI_N-\bcero_N \right]^{-1}=
\mathbf{A}_{0_{2N}}\cdot\mathbf{F}_{0_{2N}}^{-1}\; .
\end{eqnarray}

We denote by $\left[\bI_N-\bcero_N \right]$ the $N$ order identity
matrix obtained with roundoff, whose elements are characterized by
a roundoff error with absolute value $Er\leq u$ (\textit{unit
roundoff}). Thus, the $\bH$ matrix elements will be characterized
by an error whose absolute value is of order of $u$. These results
show that the MLS matrix $\bH$ converges to finite values, without
significant precision loss, when $d$ increases indefinitely.

On the other hand when $d\rightarrow 0$ we obtain immediately from
(\ref{T-d}) that $\bT \equiv \bI_{2N}$ and by substituting its
partitions of order $N$ in (\ref{H-T}) we obtain:

\begin{eqnarray}
   \bH|_{d\rightarrow0} &=& \left[\begin{array}{cc}
                              \bcero_N & \bI_N\\
                              \bI_N &  \bcero_N
                            \end{array}\right],
\end{eqnarray}

\noindent thus $\bH$ also converges in a numerically stable way when $d\rightarrow0$.

\subsection{\label{estabilidad-E}Numerical stability of the Stiffness matrix of the MSL system}

From the expressions (\ref{matriz-T}) and (\ref{matriz-E}) we derive the following relations:

\begin{eqnarray}
\label{E-T}
 \bE &=& \left[\begin{array}{cc}
-[\bT_{12}]^{-1}\cdot \bT_{11} & \;\;\;[\bT_{12}]^{-1} \\
\bT_{21}-\bT_{22}\cdot[\bT_{12}]^{-1}\cdot \bT_{11} &
\;\;\;\bT_{22}\cdot[\bT_{12}]^{-1}
         \end{array}\right]\;.
\end{eqnarray}


The Stifness matrix obtained from equation (\ref{E-T}) will be, as
the $\bH$ matrix, independent of the base of the LI solutions
chosen to build $\bT$. Because of this we consider also the ATM
coming from the expression (\ref{T-d}), which was obtained from a
base of solutions $\mathbf{F}_j(z) = \mathbf{F}_{j0} \,
e^{ik_jz}$, where $\mathbf{\Pi}_{k_N}(d)$ contains the eigenvalues
with positive imaginary part: $\Im(k_1, k_2,\ldots k_N)>0$ and
$\mathbf{\Pi}_{k_{2N}}(d)$  contains the eigenvalues with negative
imaginary part: $\Im(k_{N+1}, k_{N+2},\ldots k_{2N})<0$.

Following the same procedure employed for $\bH$ we substitute in
(\ref{E-T}) the expression of the partitions $\bT_{11}$,
$\bT_{12}$, $\bT_{21}$ and $\bT_{22}$ given in the Appendix
\ref{App-B} and calculate the limit of the partitions of $\bE$
when $d\rightarrow\infty$, to obtain:

\begin{eqnarray}
  \label{E11}\bE_{11}|_{d\rightarrow\infty} &=& \mathbf{A}_{0_{N}}\cdot\mathbf{F}_{0_{N}}^{-1}\cdot\left[\bI_N-\bcero_N \right]^{-1}=\mathbf{A}_{0_{N}}\cdot\mathbf{F}_{0_{N}}^{-1} \\
 \label{E12} \bE_{12}|_{d\rightarrow\infty} &=& \bcero_N \cdot \left[\bI_N-\bcero_N \right]^{-1}=\bcero_N \\
 \label{E21}\bE_{21}|_{d\rightarrow\infty}  &=& \left[\mathbf{A}_{0_{N}}-\mathbf{A}_{0_{2N}}\cdot\mathbf{F}_{0_{2N}}^{-1}
 \cdot\left[\bI_N-\bcero_N \right]^{-1}\cdot\mathbf{F}_{0_{N}}\right]\cdot \bcero_N \nonumber\\&&+
  \left[\mathbf{A}_{0_{2N}}-\mathbf{A}_{0_{2N}}\left[\bI_N-\bcero_N \right]^{-1}\right]
  \cdot\mathbf{\Pi}_{k_{2N}}(d)|_{d\rightarrow\infty}\cdot\gamma_{12}^{-1}=\bcero_N
  \\
  \label{E22}\bE_{22}|_{d\rightarrow\infty} &=&
  \mathbf{A}_{0_{2N}}
  \cdot\mathbf{F}_{0_{2N}}^{-1}\cdot\left[\bI_N-\bcero_N \right]^{-1}=
  \mathbf{A}_{0_{2N}}\cdot\mathbf{F}_{0_{2N}}^{-1}\;.
 \end{eqnarray}

These results show that the $\bE$ matrix also converges to finite
values without a significant precision loss when $d$ grows
indefinitely. On the other hand, when $d\rightarrow 0$, we know
that $\bT \equiv \bI_{2N}$, which means that
$\bT_{12}=\bT_{21}=\bcero_N$ and then the $\bE$ matrix is not
numerically computable (overflow) as is directly obtained from the
relations (\ref{E-T}). Let us now assume that $d$ is very small
but not enough to provoke the overflow state. From (\ref{E-T}) we
can express the partition $\bE_{21}$ in the following form:

\begin{eqnarray}
  \bE_{21} &=& -\left(\bI_N-\bT_{21}\cdot\bT_{11}^{-1}\cdot\bT_{12}\cdot\bT_{22}^{-1}\right)\cdot\bT_{22}\cdot\bT_{12}^{-1}\cdot\bT_{11},
\end{eqnarray}

\noindent then for a sufficiently small $d$ this partition will be the object of the roundoff in the first place, giving:

\begin{eqnarray}
  \bE_{21}|_{d\rightarrow 0} &=& -\left[\bI_N-\bcero_N\right]\cdot\bT_{12}^{-1}.
\end{eqnarray}

Unlike the limits given in (\ref{E11})-(\ref{E21}) the term
$\left[\bI_N-\bcero_N \right]$ subjected to the roundoff,
multiplies now a term $\bT_{12}^{-1}$ whose value can be big
enough to affect the Stiffness matrix due to the roundoff error
and then gives rise to the $\Omega d$ problem.

\section{Scattering Matrix and Coefficients Transfer Matrix for the MSL system}

In the case of the Scattering Matrix $(\bS)$ its relation with
$\bT$ is not a direct one (because there are other matrices
involved) as in the relations studied previously. A possible way
to relate $\bS$ with $\bT$ is by using the Coefficients Transfer
Matrix $(\bK)$. We need to use the direct relation $\bS$-$\bK$ and
the indirect one $\bK$-$\bT$. Being known a base of solutions
$\mathbf{F}_j(\alpha,z)$ in a domain $\alpha$ the general solution
$\mathbf{F}(\alpha,z)$ of the differential system
(\ref{Eqmaestra}) can be written as:

\begin{eqnarray}
  \mathbf{F}(\alpha,z) &=& \sum_j^{2N} a_j(\alpha)\mathbf{F}_j(\alpha,z) \;.
\end{eqnarray}

Let be  $\mathbf{a}^{+/-}\;(\alpha)$ the $N$-vector formed by the
coefficients $a_j(\alpha)$ of the amplitudes travelling to the
right/left. Then we shall denote as $\bK(\mathrm{R},\mathrm{L})$
the Coefficients Transfer Matrix transferring the ensemble of
coefficients $\mathbf{a}^{+/-}$ from domain $\mathrm{L}$ to domain
$\mathrm{R}$:

\begin{eqnarray}
 \begin{array}{|c|}
     \mathbf{a}^{+}(\mathrm{R}) \\
      \mathbf{a}^{-}(\mathrm{R})
   \end{array} &=& \bK(\mathrm{R},\mathrm{L})\cdot \begin{array}{|c|}
     \mathbf{a}^{+}(\mathrm{L}) \\
      \mathbf{a}^{-}(\mathrm{L})
   \end{array} \;.
\end{eqnarray}

The term Scattering Matrix is widely used in the literature and
can be defined in different ways. Here we shall use the definition
and notation $\bS(\mathrm{R};\mathrm{L})$ employed in
\cite{LibroUJI}:

\begin{eqnarray}
\label{S}
  \begin{array}{|c|}
     \mathbf{a}^{-}(\mathrm{L}) \\
      \mathbf{a}^{+}(\mathrm{R})
   \end{array}
   &=& \bS(\mathrm{R};\mathrm{L})\;\cdot \begin{array}{|c|}
                          \mathbf{a}^{+}(\mathrm{L}) \\
                          \mathbf{a}^{-}(\mathrm{R})
                       \end{array}\;.
\end{eqnarray}

From these definitions we obtain a direct relation between $\bS$ and $\bK$:

\begin{eqnarray}
\label{S-K}
  \bS &=& \left[\begin{array}{cc}
            -[\bK_{22}]^{-1}\cdot\bK_{21} & \;\;\;[\bK_{22}]^{-1} \\
            \bK_{11}-\bK_{12}\cdot[\bK_{22}]^{-1}\cdot\bK_{21} & \;\;\;\bK_{12}\cdot [\bK_{22}]^{-1}
          \end{array}\right] \;.
\end{eqnarray}

By taking into account that between the domains $\mathrm{R}$ and
$\mathrm{L}$ there is an intermediate region $\mathrm{M}$ (can be
a single or a multiple layer) described by a $\bT$ matrix, it is
possible to obtain \cite{LibroUJI}:

\begin{eqnarray}\label{K-T}
  \bK(\mathrm{R},\mathrm{L}) &=& [\bQ(\mathrm{R}:z_r)]^{-1}\cdot \bT(z_r,z_\ell)\cdot \bQ(\mathrm{L}:z_\ell)\;,
\end{eqnarray}

\noindent where $z_{r/\ell}$ are the coordinates of the interfaces
matching the intermediate region $\mathrm{M}$ with the external
domains $\mathrm{R}$ (to the right)/$\mathrm{L}$ (to the left).
The matrix $\bQ(z)$ for an arbitrary domain is given in
(\ref{Q-ConsT}). Expression (\ref{K-T}) shows clearly that the
matrix $\bK$ and consequently $\bS$ depends on the base of LI
solutions chosen to build the matrix $\bQ$. It is a common
practice to choose a reduced base in $z_{r/\ell}$, that is a base
tending to unity in $z_{r/\ell}$.

\subsection{\label{estabilidad-S}Analysis of the numerical stability of the Scattering Matrix (SM)}

In the first place we substitute in (\ref{K-T}) the expressions
(\ref{particiones-T}) giving the $\bT$ partitions when
$d\rightarrow\infty$. Now we substitute in (\ref{S-K}) the
expressions obtained for the partitions of $\bK$ and obtain:

\begin{eqnarray}
 \bS_{11}|_{d\rightarrow\infty} &=& -\left[\gamma_{12}^{-1}\cdot\left(\bI_N+\bcero_N \right)\cdot\bQ(z_\ell)_{12}+\gamma_{22}^{-1}\cdot\left(\bI_N+\bcero_N \right)\cdot\bQ(z_\ell)_{22}\right]^{-1}\nonumber\\&&\cdot\left[\gamma_{12}^{-1}\cdot\left(\bI_N+\bcero_N \right)\cdot\bQ(z_\ell)_{11}+\gamma_{22}^{-1}\cdot\left(\bI_N+\bcero_N \right)\cdot\bQ(z_\ell)_{21}\right]\;;\\
  \bS_{12}|_{d\rightarrow\infty} &=& \left[\gamma_{12}^{-1}\cdot\left(\bI_N+\bcero_N \right)\cdot\bQ(z_\ell)_{12}+\gamma_{22}^{-1}\cdot\left(\bI_N+\bcero_N \right)\cdot\bQ(z_\ell)_{22}\right]^{-1}\nonumber\\&&\cdot\mathbf{\Pi}_{k_{2N}}(-d)|_{d\rightarrow\infty}\cdot\left[(\bQ(z_r)^{-1})_{21}\cdot\mathbf{F}_{0_{2N}}+(\bQ(z_r)^{-1})_{22}\cdot\mathbf{A}_{0_{2N}}\right]^{-1}\nonumber\\&&=\bcero_N  \;;\\
 \bS_{21}|_{d\rightarrow\infty} &=& \left[(\bQ(z_r)^{-1})_{11}\cdot\mathbf{F}_{0_{2N}}+(\bQ(z_r)^{-1})_{12}\cdot\mathbf{A}_{0_{2N}}\right]\cdot\mathbf{\Pi}_{k_{2N}}(d)|_{d\rightarrow\infty}\nonumber\\&&\cdot \left[\gamma_{12}^{-1}\cdot\left(\bI_N+\bcero_N \right)\cdot\bQ(z_\ell)_{11}+\gamma_{22}^{-1}\cdot\left(\bI_N+\bcero_N \right)\cdot\bQ(z_\ell)_{21}\right]\nonumber\\&& -\mathrm{Identical}=\bcero_N\;;\\
 \bS_{22}|_{d\rightarrow\infty} &=&  \left[(\bQ(z_r)^{-1})_{11}\cdot\mathbf{F}_{0_{2N}}+(\bQ(z_r)^{-1})_{12}\cdot\mathbf{A}_{0_{2N}}\right]\nonumber\\&&\cdot\left[(\bQ(z_r)^{-1})_{21}\cdot\mathbf{F}_{0_{2N}}+(\bQ(z_r)^{-1})_{22}\cdot\mathbf{A}_{0_{2N}}\right]^{-1}.
\end{eqnarray}

We have used the notation $\bQ(z_r)$ instead of $\bQ(\mathrm{R}:z_r)$ and $\bQ(z_\ell)$ instead of $\bQ(\mathrm{L}:z_\ell)$ to simplify the expressions. The term $\left[\bI_N+\bcero_N \right]$ is an identity matrix of order $N$ obtained by roundoff whose elements have a roundoff error with absolute value $Er\leq u$ (\textit{unit roundoff}). Because of this the matrix elements of $\bS$ will have an error with absolute value of the order of $u$. These results show that the $\bS$ matrix of a MSL system converges also to finite values without significant precision loss when $d$ increases indefinitely.

On the other hand, if the intermediate region $M$ thickness is nil (no $M$ region, $z_\ell=z_r=z_s$) we obtain from (\ref{K-T}) that $\bK(\mathrm{R},\mathrm{L})= [\bQ(\mathrm{R}:z_s)]^{-1}\cdot\bQ(\mathrm{L}:z_s)$ and we can obtain $\bS$ without trouble. This means that the $\bS$ matrix of the MSL can avoid the $\Omega d$ problem and converge in a stable numerical way when $d\rightarrow0$.

\subsection{\label{ReglasdeC}Composition rules}

The hybrid matrix $\bH^{(m)}$ relating the field and the linear
form in the positions $z_{m-1}$ and $z_r$ of the sketch shown in
Figure \ref{Figura01} can be described as the hybrid matrix of the
structure formed by the layers $m$, $m+1$,\ldots ,$\mu$:

\begin{eqnarray}
\label{matriz-H-2}
  \begin{array}{|c|}
    \mathbf{F}(m:z_{m-1}) \\
     \mathbf{A} (\mu:z_r)
   \end{array}
   &=& \bH^{(m)}\;\cdot \begin{array}{|c|}
                           \mathbf{A}(m:z_{m-1}) \\
                           \mathbf{F}(\mu:z_r)
                         \end{array}\;.
\end{eqnarray}

We use here the supraindex $m$ among parentheses to denote the
hybrid matrix of the structure being considered in a similar way
to that employed in Ref. \cite{2006-ELTan-ASA}. The partitions of
the matrix given in (\ref{matriz-H-2}) can be expressed in terms
of the $\bH^{(m+1)}$ matrix partitions corresponding to the
structure including the layers from $m+1$ to $\mu$ and of the
matrix $\bH^{m}$ given by (\ref{matriz-H-1}) relating the field
and the linear form in the layer $m$ borders $z_0=z_{m-1}$ and
$z=z_m$. We must take into account the continuity of the field and
the associated linear form in $z_{m}$,
$\mathbf{F/A}(m+1:z_{m})=\mathbf{F/A}(m:z_{m})$. Then we obtain
the following composition rule:

\begin{eqnarray}\label{regla-composicion-H}
  \bH^{(m)}_{11} &=& \bH^m_{11}+\bH^m_{12}\cdot\bH^{(m+1)}_{11}\cdot \left[\bI_N-\bH^m_{22}\cdot\bH^{(m+1)}_{11}\right]^{-1}\cdot\bH^m_{21}\; \nonumber\\
  \bH^{(m)}_{12} &=& \bH^m_{12}\cdot\left[\bI_N+\bH^{(m+1)}_{11}\cdot \left[\bI_N-\bH^m_{22}\cdot\bH^{(m+1)}_{11}\right]^{-1}\cdot\bH^m_{22}\right]\cdot\bH^{(m+1)}_{12};\nonumber\\
  \bH^{(m)}_{21} &=& \bH^{(m+1)}_{21}\cdot\left[\bI-\bH^m_{22}\cdot\bH^{(m+1)}_{11}\right]^{-1}\cdot\bH^m_{21};\nonumber \\
  \bH^{(m)}_{22} &=& \bH^{(m+1)}_{22}+\bH^{(m+1)}_{21}\cdot\left[\bI_N-\bH^m_{22}\cdot\bH^{(m+1)}_{11}\right]^{-1}\cdot\bH^m_{22}\cdot\bH^{(m+1)}_{12}.
\end{eqnarray}

In the same way the Stiffness matrix relating the field and the
linear form in the positions $z_{m-1}$ and $z_r$ can be described
as the Stiffness matrix of the structure formed by the layers $m$,
$m+1$, \ldots, $\mu$:

\begin{eqnarray}
\label{matriz-E-2}
  \begin{array}{|c|}
    \mathbf{A}(m:z_{m-1}) \\
     \mathbf{A}(\mu:z_r)
   \end{array}
   &=& \bE^{(m)}\;\cdot \begin{array}{|c|}
                           \mathbf{F}(m:z_{m-1}) \\
                           \mathbf{F}(\mu:z_r)
                         \end{array}\;,
\end{eqnarray}

\noindent and its composition rule in terms of $\bE^{(m+1)}$ and $\bE^{m}$ is given by:

\begin{eqnarray}\label{regla-composicion-E}
 \bE^{(m)}_{11} &=&  \bE^{m}_{11}+\bE^{m}_{12}\cdot \left[\bE^{(m+1)}_{11}-\bE^{m}_{22}\right]^{-1}\cdot\bE^{m}_{21};\nonumber\\
  \bE^{(m)}_{12} &=& -\bE^{m}_{12}\cdot \left[\bE^{(m+1)}_{11}-\bE^{m}_{22}\right]^{-1}\cdot\bE^{(m+1)}_{12};\nonumber\\
  \bE^{(m)}_{21} &=& \bE^{(m+1)}_{21}\cdot \left[\bE^{(m+1)}_{11}-\bE^{m}_{22}\right]^{-1}\cdot\bE^{m}_{21}; \nonumber\\
  \bE^{(m)}_{22} &=& \bE^{(m+1)}_{22}-\bE^{(m+1)}_{21}\cdot\left[\bE^{(m+1)}_{11}-\bE^{m}_{22}\right]^{-1}\cdot\bE^{(m+1)}_{12}
\end{eqnarray}

Analogously, the composition rule for the Scattering matrix
$\bS(\mathrm{R};\mathrm{m})$, can be expressed in terms of the
matrices $\bS(\mathrm{R};\mathrm{m+1})$ and
$\bS(\mathrm{m+1};\mathrm{m})$, each one defined in agreement with
(\ref{S}) for the interfaces placed between the domains
$\mathrm{m}$ and $\mathrm{R}$, between $\mathrm{m+1}$ and
$\mathrm{R}$ and between $\mathrm{m}$ and $\mathrm{m+1}$,
respectively. This composition rule can be expressed by means of
the product denoted by $\bigocoasterisk$, in the form:

\begin{eqnarray}
  \bS(\mathrm{R};\mathrm{m}) &=& \bS(\mathrm{R};\mathrm{m+1}) \bigocoasterisk  \bS(\mathrm{m+1};\mathrm{m})\;.
\end{eqnarray}

Given three matrices $\mathbf{Z}$, $\mathbf{Y}$ and $\mathbf{X}$
of order 2$N$ subdivided in their $N\times N$ partitions the
product $\bigocoasterisk$ expressing
$\mathbf{Z}=\mathbf{Y}\bigocoasterisk \mathbf{X}$ is defined in
\cite{LibroUJI} by means of the composition rule:

\begin{eqnarray}\label{regla-composicion-S}
   \mathbf{Z}_{11}&=&\mathbf{X}_{11}+\mathbf{X}_{12}\cdot\mathbf{Y}_{11}\cdot\left[\bI_N-\mathbf{X}_{22}\cdot\mathbf{Y}_{11}\right]^{-1}\cdot\mathbf{X}_{21}; \nonumber\\
   \mathbf{Z}_{12} &=& \mathbf{X}_{12}\cdot\mathbf{Y}_{12}+\mathbf{X}_{12}\cdot\mathbf{Y}_{11}\cdot\left[\bI_N-\mathbf{X}_{22}\cdot\mathbf{Y}_{11}\right]^{-1}\cdot\mathbf{X}_{22}\cdot\mathbf{Y}_{12};\nonumber \\
   \mathbf{Z}_{21} &=& \mathbf{Y}_{21}\cdot\left[\bI_N-\mathbf{X}_{22}\cdot\mathbf{Y}_{11}\right]^{-1}\cdot\mathbf{X}_{21};\nonumber \\
   \mathbf{Z}_{22} &=& \mathbf{Y}_{22}+\mathbf{Y}_{21}\cdot\left[\bI_N-\mathbf{X}_{22}\cdot\mathbf{Y}_{11}\right]^{-1}\cdot\mathbf{X}_{22}\cdot\mathbf{Y}_{12}.
\end{eqnarray}

We must note that the composition rules
(\ref{regla-composicion-H}) and (\ref{regla-composicion-S})
include the inverses
$\left[\bI_N-\bH^m_{22}\cdot\bH^{(m+1)}_{11}\right]^{-1}$ and
$\left[\bI_N-\bS_{22}(\mathrm{m+1};\mathrm{m})\cdot\bS_{11}(\mathrm{R};\mathrm{m+1})\right]^{-1}$
respectively, which are regular even when the thickness of the
layer or of the multilayer goes to infinity or to zero. The
composition rule (\ref{regla-composicion-E}) includes the term
$\left[\bE^{(m+1)}_{11}-\bE^{m}_{22}\right]^{-1}$, which is
regular when the thickness of the layer or of the multilayer goes
to infinity. For very small thicknesses this composition rule will
lead to the accumulation of the roundoff errors.

\section{\label{aplicaciones}General formulation of some typical boundary problems. Numerical examples}

The boundary problems can be formulated in terms of the $\bT$,
hybrid, scattering or stiffness matrices. We consider a system
formed by three domains $\mathrm{L}-\mathrm{M}-\mathrm{R}$. The
internal domain $\mathrm{M}$ can be formed by one or several
homogeneous layers in whose case the matrix of the structure
$\mathrm{M}$ must be obtained through composition rules (see
Section \ref{ReglasdeC}). In the external domains $\mathrm{L}$ and
$\mathrm{R}$ we can have different media and even the vacuum.
Depending on the problem under study we shall employ different
boundary conditions at the interface $\mathrm{L}|\mathrm{M}$ with
coordinate $z_\ell$ and at $\mathrm{M}|\mathrm{R}$ with coordinate
$z_r$. We denote by $\mathbf{F}(\ell/r)$, $\mathbf{A}(\ell/r)$ the
field and the associated linear form at the coordinate
$z_{\ell}/z_r$. In all the cases here considered we avoid to use
submatrices which can exhibit numerical instabilities when
$d\rightarrow\infty$, as it happens for $\bH_{12}^{-1}$ or
$\bE_{12}^{-1}$.

\subsection{Escape problem}

We shall study the escape problem in a system formed by three
media $\mathrm{L}-\mathrm{M}-\mathrm{R}$ having full matching
conditions (FMC) at the interface $\mathrm{L}|\mathrm{M}$ with
coordinate $z_\ell$ and at the interface $\mathrm{M}|\mathrm{R}$
with coordinate $z_r$. In the scape problem we shall have only
outgoing waves in $\mathrm{M}$. Applying the continuity conditions
at the interface we can write:

\begin{eqnarray}\label{H-r-l-escape}
  \begin{array}{|c|}
     \mathbf{F}(\ell)^{-} \\
      \mathbf{A}(r)^{+}
   \end{array}
   &=& \bH(r,\ell)\;\cdot \begin{array}{|c|}
                          \mathbf{A}(\ell)^{-} \\
                          \mathbf{F}(r)^{+}
                       \end{array}\;.
\end{eqnarray}

The superindex $\pm$ denote the vectors related with the wave
travelling in $\mathrm{R}/\mathrm{L}$ towards the right/left. From
the two matrix equations coming from (\ref{H-r-l-escape}) we can
write:

\begin{eqnarray}\label{Sistema-1}
  \bcero &=& \left(\begin{array}{cccc}
               -\bI_N &\; \bH_{11}(r,\ell) &\; \bH_{12}(r,\ell) &\; \bcero_N \\
               \bcero_N &\; \bH_{21}(r,\ell) &\; \bH_{22}(r,\ell) &\; -\bI_N
             \end{array}\right)\cdot\left|\begin{array}{c}
                                  \mathbf{F}(\ell)^{-} \\
                                  \mathbf{A}(\ell)^{-} \\
                                  \mathbf{F}(r)^{+} \\
                                  \mathbf{A}(r)^{+}
                               \end{array}\right|\;.
\end{eqnarray}

We can express the vectors appearing in the right-hand side of (\ref{Sistema-1}) in the form:

\begin{eqnarray}\label{LI-izquierda}
\left| \begin{array}{c}
   \mathbf{F}(\ell)^{-} \\
    \mathbf{A}(\ell)^{-}
 \end{array}\right|
  &=& \left[\begin{array}{ccc}
                    \mathbf{F}_1(\ell)^{-}  & \ldots &  \mathbf{F}_{N}(\ell)^{-} \\
                    \mathbf{A}_1 (\ell)^{-} & \ldots & \mathbf{A}_{N} (\ell)^{-}
                  \end{array}\right]\cdot \left|\begin{array}{c}
                                            a_1(\mathrm{L})^{-} \\
                                            \vdots \\
                                            a_N(\mathrm{L})^{-}
                                          \end{array}\right|
                  =\mathbf{LI}(\ell)^{-}\cdot\mathbf{a}(\mathrm{L})^{-}\;,
\end{eqnarray}

\begin{eqnarray}\label{LI-derecha}
\left| \begin{array}{c}
   \mathbf{F}(r)^{+} \\
    \mathbf{A}(r)^{+}
 \end{array}\right|
  &=& \left[\begin{array}{cccc}
                    \mathbf{F}_1(r)^{+}  & \ldots &  \mathbf{F}_{N}(r)^{+} \\
                    \mathbf{A}_1 (r)^{+} & \ldots & \mathbf{A}_{N} (r)^{+}
                  \end{array}\right]\cdot \left|\begin{array}{c}
                                            a_1(\mathrm{R})^{+} \\
                                            \vdots \\
                                            a_N(\mathrm{R})^{+}
                                          \end{array}\right|=\mathbf{LI}(r)^{+}\cdot\mathbf{a}(\mathrm{R})^{+}\;,
\end{eqnarray}

\noindent where $\mathbf{F}_j(\ell)^{-}$ are LI solutions
belonging to the $\mathrm{L}$ domain, evaluated at $z_\ell$ and
$\mathbf{F}_j(r)^{+}$ are LI solutions belonging to the
$\mathrm{R}$ domain, evaluated at $z_r$. The $N$-vector
$\mathbf{a}(\mathrm{R})^{+}$ is formed by the coefficients
$a_j(\mathrm{R})^{+}$ from those waves travelling to the right at
$\mathrm{R}$ and $\mathbf{a}(\mathrm{L})^{-}$ by the coefficients
$a_j(\mathrm{L})^{-}$ from those waves travelling to the left at
$\mathrm{L}$.

Then by using (\ref{LI-izquierda}) and (\ref{LI-derecha}) we
transform (\ref{Sistema-1}) into the secular system:

\begin{eqnarray}
  \bcero &=& \left(\begin{array}{cc}
               \mathbf{Ms}_{11} & \mathbf{Ms}_{12} \\
               \mathbf{Ms}_{21} & \mathbf{Ms}_{22}
             \end{array}\right)\cdot \left|\begin{array}{c}
                                       \mathbf{a}(\mathrm{L})^{-} \\
                                       \mathbf{a}(\mathrm{R})^{+}
                                     \end{array}\right|\;;
\end{eqnarray}

\begin{eqnarray}
\label{Ms-H-11}  \mathbf{Ms}_{11} &=& \left[\begin{array}{cc}
                         -\bI_N &\;\; \bH_{11}(r,\ell)
                       \end{array}\right]\cdot\mathbf{LI}(\ell)^{-}, \\
\label{Ms-H-12} \mathbf{Ms}_{12} &=& \left[\begin{array}{cc}
                         \bH_{12}(r,\ell) &\;\; \bcero_N
                       \end{array}\right]\cdot\mathbf{LI}(r)^{+}, \\
 \label{Ms-H-211} \mathbf{Ms}_{21} &=& \left[\begin{array}{cc}
                         \bcero_N &\;\; \bH_{21}(r,\ell)
                       \end{array}\right]\cdot\mathbf{LI}(\ell)^{-}, \\
 \label{Ms-H-22}  \mathbf{Ms}_{22} &=& \left[\begin{array}{cc}
                         \bH_{22}(r,\ell) &\;\; -\bI_N
                       \end{array}\right]\cdot\mathbf{LI}(r)^{+}\;.
\end{eqnarray}

The problem eigenvalues are obtained from the secular equation $\Det[\mathbf{Ms}]=0$.

In terms of the Stiffness matrix we have:

\begin{eqnarray}
  \mathbf{Ms}_{11} &=& \left[\begin{array}{cc}
                        \bE_{11}(r,\ell) &\;\; -\bI_N
                       \end{array}\right]\cdot\mathbf{LI}(\ell)^{-}, \\
 \mathbf{Ms}_{12} &=& \left[\begin{array}{cc}
                         \bE_{12}(r,\ell) &\;\; \bcero_N
                       \end{array}\right]\cdot\mathbf{LI}(r)^{+}, \\
  \mathbf{Ms}_{21} &=& \left[\begin{array}{cc}
                         \bE_{21}(r,\ell) &\;\; \bcero_N
                       \end{array}\right]\cdot\mathbf{LI}(\ell)^{-}, \\
   \mathbf{Ms}_{22} &=& \left[\begin{array}{cc}
                         \bE_{22}(r,\ell) &\;\; -\bI_N
                       \end{array}\right]\cdot\mathbf{LI}(r)^{+}\;.
\end{eqnarray}

As a numerical example we use the secular equation in terms of the
hybrid matrix $\bH(r,\ell)$ to obtain the velocities of shear
horizontal (SH) acoustic waves in $\mu$ piezoelectric multilayers
systems. These curves were obtained in Ref. \cite{2010-HCalas-JAP}
by using the singular value decomposition (SVD) method together
with a variant of the Global Matrix Method (GMM) as an alternative
technique to avoid the numerical instabilities found by the
authors.

The piezoelectric systems studied there, are formed by two
different materials, A (PZT4) and B (PZT5A), and have different
layer configurations: $n=3$ (ABA), $n=5$ (ABABA), $n=7$ (ABABABA)
and $n=9$ (ABABABABA). All these systems have the
$\mathrm{L}-\mathrm{M}-\mathrm{R}$ structure, with
$\mathrm{L}=\mathrm{R}=A$. The external domains are semi-infinite
and to obtain confined modes it was assumed that there are no
ingoing waves in the inner region $\mathrm{M}$, whereas the
outgoing waves are evanescent. It is then clear that this problem
can be studied as a particular case of the scape problem
considered in this section. In order to get evanescent waves the
eigenvalues $k_j$ appearing in the exponential terms of these
waves were assumed to be pure imaginary.

Except for $n=3$, the hybrid matrix $\bH(r,\ell)$ in the inner
region $\mathrm{M}$ was obtained by means of the composition rule
(\ref{regla-composicion-H}). To solve this problem it was
necessary to transform the original system of two equations of
motion \cite{2010-HCalas-JAP} in a matrix Sturm-Liouville system
(\ref{Eqmaestra}) with $N$=2. In this problem $\mathbf{F}(z)$ has
two components, the transverse displacement $u$ and the electric
potential $\phi$. The $z$ axis is oriented in the direction normal
to the multilayer interfaces in such a way that it coincides with
the $y$ axis in the scheme of Figure 1 in Ref.
\cite{2010-HCalas-JAP}. The $x$ axis coincides in both cases.

The quadratic eigenvalues problem solution (QEP, Section
\ref{QEP}) for one layer is:

\begin{eqnarray}
  k_1 &=& -i \kappa_x=-i \frac{\omega}{v_s} \\
  k_2 &=& -k_1 \\
 \label{k3} k_3 &=& -\sqrt{-\kappa_x^2+ \omega^2 \frac{\rho}{(c_{44}+
 \displaystyle{\frac{e^2_{15}}{\epsilon_{11}})}}}=- i \omega \sqrt{\frac{1}{v_s^2}-\frac{1}{v^2}}\\
  k_4 &=& - k_3\;,
\end{eqnarray}

\noindent where $v_s$ is the velocity of the surface wave we are
studying, whereas
$v=\sqrt{(c_{44}+\displaystyle{\frac{e^2_{15}}{\epsilon_{11}}})/\rho}$
is the SH wave velocity. The material parameters of the layer
needed for this study are the mass density $\rho$, the elastic
constant $c_{44}$, the piezoelectric constant $e_{15}$ and the
dielectric constant $\epsilon_{11}$. The eigenfunctions can be
chosen in the form:

\begin{eqnarray}
\mathbf{F}_{j}(z) &=& \mathbf{F}_{j0}\;e^{ik_j(z-z_0)} = \left(\begin{array}{c}
                        0 \\
                        1
                      \end{array}\right)\;e^{ik_j(z-z_0)}, \;\; j=1,2
\end{eqnarray}

\noindent and:

\begin{eqnarray}
   \mathbf{F}_{j}(z) &=& \mathbf{F}_{j0}\;e^{ik_j(z-z_0)} = \left(\begin{array}{c}
                        1 \\
                        e_{15}/\epsilon_{11}
                      \end{array}\right)\;e^{ik_j(z-z_0)}\;\; j=3,4.
\end{eqnarray}

$k_3$ and $k_4$ must be pure imaginary to obtain evanescent
outgoing waves. As these waves travel in material A (PZT4) layers
the expression (\ref{k3}) shows that this happens for $v_s<v_A$.
It is also possible to obtain confined modes when there are layers
in the domain $\mathrm{M}$ with $k_3$ and $k_4$ real. This is only
possible in material B (PZT5A) layers when $v_B<v_s$.

The hybrid matrix of an independent layer was obtained by a method
analogous to that employed in \cite{LibroUJI} to get the
expression (\ref{T-Q}). For the $\bH$ matrix we have:

\begin{eqnarray}
  \bH (z,z_0)&=& \mathbf{U}^{FA}(z,z_0)\cdot\left[\mathbf{U}^{AF}(z,z_0)\right]^{-1}.
\end{eqnarray}

\begin{eqnarray}
  \mathbf{U}^{FA}(z,z_0) &=& \left[\begin{array}{cccc}
                               \mathbf{F}_{1}(z_0) & \mathbf{F}_{2}(z_0) & \ldots & \mathbf{F}_{2N}(z_0) \\
                               \mathbf{A}_{1}(z) &  \mathbf{A}_{2}(z) &  \ldots &  \mathbf{A}_{2N}(z)
                             \end{array}\right];
\end{eqnarray}

\begin{eqnarray}
  \mathbf{U}^{AF}(z,z_0) &=& \left[\begin{array}{cccc}
                               \mathbf{A}_{1}(z_0) & \mathbf{A}_{2}(z_0) & \ldots & \mathbf{A}_{2N}(z_0) \\
                               \mathbf{F}_{1}(z) &  \mathbf{F}_{2}(z) &  \ldots &  \mathbf{F}_{2N}(z)
                             \end{array}\right].
\end{eqnarray}

The secular matrix $\mathbf{Ms}$ was obtained from the expressions
(\ref{Ms-H-11}-\ref{Ms-H-22}) and then we obtained the values of
the surface wave velocities zeroing the secular determinant at
different frequency values. Table \ref{Tabla1} shows the values
obtained in our calculation, those obtained in
\cite{2010-HCalas-JAP} together with the corresponding
frequencies. The values in \cite{2010-HCalas-JAP} were obtained by
using the (SVD) method and a Global Matrix of order $4(N_{L}-1)
\times 4(N_{L}-1)$, $N_{L}$ being the number of layers in the
structure. Thus for $N_{L}=3$ the matrix would be $(8 \times 8)$,
whereas for $N_{L}=9$ the matrix would be $(32 \times 32)$. The
hybrid matrix employed in our calculations is of order
$(4\times4)$. The good agreement of both sets of velocity values
shows the capability of the hybrid matrix method to avoid the
$\Omega d$ problem with lower computational and formal
requirements when compared with the Global Matrix method.

\begin{figure}
\begin{center}
\includegraphics[width=4.0in,height=3.0in]
{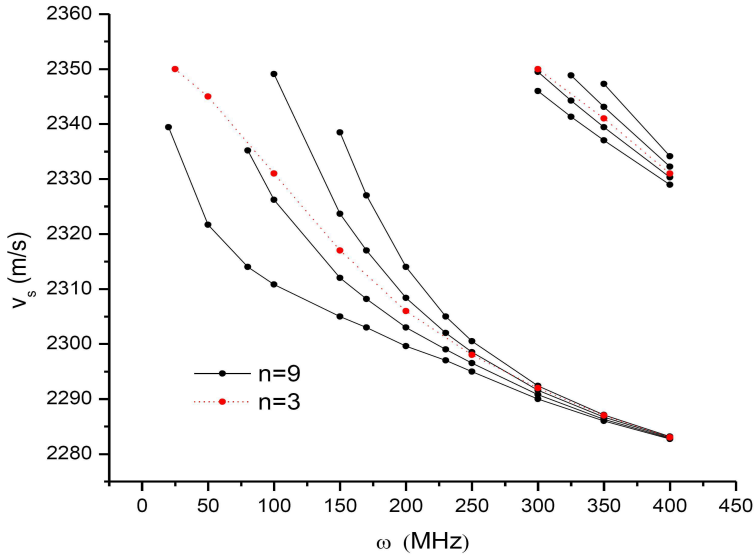}\\
\caption{\label{Fig1} Surface wave velocity values for different frequency values of the $n=3$ and $n=9$ systems}
\end{center}
\end{figure}

Figure \ref{Fig1} shows the values of the surface wave velocity
for the corresponding frequency values for the three and nine
layer systems coming from our calculations. We can observe two
bands, the first of the even modes and the first of the odd modes,
together with the convergence of the modes of the system $N$=9
towards those of the system $N$=3 when the frequency is increased.
This behaviour is present in the curves given in
\cite{2010-HCalas-JAP}.

\begin{table}
\begin{center}
\begin{tabular}{|c|c|c|c|}\hline \hline
   No. of    & \;$\omega$\;  &  \textbf{MG} and \textbf{SVD}        &   $\bH s$\\
      layers       &  (MHz)         & \;vs\;(m/s)        &\;vs\;(m/s)  \\\hline
   3              & 123.1          & 2324                & 2324.08      \\
                  & 357.1          & 2286                & 2285.94      \\\hline
  5               & 123.1          & \;2313.6/\;2340.5\;   & 2313.9/\;2340.8\\
                  & 279.1          & 2292.6/\;2294.5       & 2292.5/\;2294.7\\
                  & 318.1          & 2343.1/\;2350.7       & 2344.1/\;2351\\
                  & 396.1          & 2330.3/\;2333.6       & 2330.3/\;2333.8\\\hline
  9               & 20             & 2339                & 2339.4\\
                  & 80             & 2314/\;2335           & 2314/\;2335.2\\\hline \hline
\end{tabular}
\caption{\label{Tabla1} Comparison between the surface wave
velocity values for different frequency values obtained by two
different theoretical methods: (\textbf{GM}) \textit{Global Matrix
Method} and (\textbf{SVD}) \textit{Singular Value Decomposition
method}.  ($\bH s$)  \textit{Hybrid compliance-stiffness Matrix
Method}.}
\end{center}
\end{table}

\subsection{Periodic systems}

Let us consider a periodic system along the $z$ direction with
arbitrary period $d$. This could be a periodic bulk crystal or a
superlattice. The matrices $\bH(z+d, z)$, $\bE(z+d, z)$ transfer
along a given period. The Bloch-Floquet conditions are satisfied
for both $\mathbf{F}(z)$ and $\mathbf{A}(z)$, in such a way that
$\mathbf{F}(z+d) = \mathbf{F}(z)\cdot e^{iq d}$ and
$\mathbf{A}(z+d) = \mathbf{A}(z)\cdot e^{iq d}$.

We can write this in terms of the hybrid matrix as:

\begin{eqnarray}
\label{H-SPeriodico}
  \begin{array}{|c|}
     \mathbf{F}(z) \\
     \mathbf{A}(z+d)
   \end{array}
   &=& \bH(z+d, z)\;\cdot \begin{array}{|c|}
                           \mathbf{A}(z) \\
                           \mathbf{F}(z+d)
                         \end{array}\;.
\end{eqnarray}

The Bloch-Floquet conditions for $\mathbf{F}(z)$ and $\mathbf{A}(z)$ in (\ref{H-SPeriodico}) lead to:

\begin{eqnarray}
 \label{Ec-1} \mathbf{A}(z) &=& \bH_{11}^{-1}\cdot \left[\bI-\bH_{12}e^{iq d} \right]\cdot\mathbf{F}(z) \\
 \label{Ec-2} \mathbf{A}(z) &=& \left[\bI-\bH_{21}e^{-iq d}\right]\cdot \bH_{22}\cdot \mathbf{F}(z);.
\end{eqnarray}

We write $\bH$ instead of $\bH(z+d, z)$ to simplify. The secular
system is obtained from expressions equations (5.24) and (5.25):

\begin{eqnarray}
\left\{\left[\bI-\bH_{21}e^{-iq d}\right]\cdot\bH_{22}-\bH_{11}^{-1}\cdot\left[\bI-\bH_{12}e^{iq d} \right]\right\}\cdot\mathbf{F}(z)  &=& \bcero_N\;.
\end{eqnarray}

\noindent It will have nontrivial solutions if:

\begin{eqnarray}
\label{r-dispersionH-N}
  \Det \left\{\left[\bI-\bH_{21}e^{-iq d}\right]\cdot\bH_{22}-\bH_{11}^{-1}\cdot\left[\bI-\bH_{12}e^{iq d} \right]\right\} &=& 0.
\end{eqnarray}

This equation gives a dispersion relation in terms of the $\bH$ matrix elements for any $N$.

Following the same procedure with $\bE(z+d, z)$ we obtain the following secular system:

\begin{eqnarray}
  \left\{\left[\bE_{11}+\bE_{12}e^{iqd} \right]-\left[\bE_{21}e^{-iqd}+\bE_{22} \right]\right\}\cdot \mathbf{F}(z) &=& \bcero_N\;,
\end{eqnarray}

\noindent and the dispersion relation:

\begin{eqnarray}
\label{r-dispersionE-N}
  \Det  \left\{\left[\bE_{11}+\bE_{12}e^{iqd} \right]-\left[\bE_{21}e^{-iqd}+\bE_{22} \right]\right\} &=& 0
\end{eqnarray}

We assume that in our periodic system the inner domain
$\mathrm{M}$ (containing one or several homogeneous layers)
coincides with the period $d$. Now we shall pose the problem in
terms of the Scattering matrix $\bS(\mathrm{R};\; \mathrm{L})$.

For the external domains $\mathrm{L}$ and $\mathrm{R}$ we have:

\begin{eqnarray}
\label{FA-Q-l}
  \begin{array}{|c|}
    \mathbf{F}({\rm L}:z_\ell) \\
    \mathbf{A} ({\rm L}:z_\ell)
  \end{array}
   &=& \bQ({\rm L}:z_\ell)\cdot \begin{array}{|c|}
                           \mathrm{ a}^{+}({\rm L}) \\
                           \mathrm{a}^{-}({\rm L})
                          \end{array}.
\end{eqnarray}

\begin{eqnarray}
\label{FA-Q-r}
  \begin{array}{|c|}
    \mathbf{F}({\rm R}:z_r) \\
    \mathbf{A} ({\rm R}:z_r)
  \end{array}
   &=& \bQ({\rm R}:z_r)\cdot \begin{array}{|c|}
                           \mathrm{ a}^{+}({\rm R}) \\
                           \mathrm{a}^{-}({\rm R})
                          \end{array}.
\end{eqnarray}

From now on we shall employ $\bQ{\rm L}$ instead of  $\bQ({\rm
L}:z_\ell)$ and $\bQ{\rm R}$ instead of $\bQ({\rm R}:z_r)$ to
simplify the notation. Usually a reduced base in $z_\ell$ is
employed to obtain $\bQ{\rm L}$ and a reduced base in $z_r$ is
used to obtain $\bQ{\rm R}$. From these matrices we can obtain the
matrix $\bK(\mathrm{R},\mathrm{L})$ by means of (\ref{K-T}) and
then from (\ref{S-K}) we can obtain $\bS(\mathrm{R};\;
\mathrm{L})$.

From the Bloch-Floquet condition we obtain:

{\large
\begin{eqnarray}
\label{Condicion-Bloch-S}
  \begin{array}{|c|}
    {\Bbb F}({\rm R}:z_r) \\
    \Bbb A ({\rm R}:z_r)
  \end{array}
   &=& \begin{array}{|c|}
         {\Bbb F}({\rm L}:z_\ell) \\
         \Bbb A ({\rm L}:z_\ell)
       \end{array}\;\; e^{iq d}
\end{eqnarray}}

Combining (\ref{FA-Q-l}), (\ref{FA-Q-r}) and
(\ref{Condicion-Bloch-S}) with the expression (\ref{S}) defining
the Scattering matrix we can write the following expressions:

\begin{eqnarray}
  \mathbf{a}^{+}({\rm L}) &=& \left[\bQ{\rm R}_{11}\cdot\bS_{21}-\bQ{\rm L}_{11}\;e^{iq d}-\bQ{\rm L}_{12}\cdot\bS_{11}\;e^{iq d}\right]^{-1} \cdot\nonumber\\&& \left[\bQ{\rm L}_{12}\cdot\bS_{12}\;e^{iq d}-\bQ{\rm R}_{11}\cdot\bS_{22}-\bQ{\rm R}_{12}\right]\cdot \mathbf{a}^{-}({\rm R}).
\end{eqnarray}

\begin{eqnarray}
  \mathbf{a}^{+}({\rm L}) &=& \left[\bQ{\rm R}_{21}\cdot\bS_{21}-\bQ{\rm L}_{21}\;e^{iq d}-\bQ{\rm L}_{22}\cdot\bS_{11}\;e^{iq d}\right]^{-1} \cdot \nonumber\\&&\left[\bQ{\rm L}_{22}\cdot\bS_{12}\;e^{iq d}-\bQ{\rm R}_{21}\cdot\bS_{22}-\bQ{\rm R}_{22}\right]\cdot \mathbf{a}^{-}({\rm R}).
\end{eqnarray}

Subtracting these equations we arrive to the secular system:

\begin{eqnarray}
  \bcero_N &=& \mathbf{Ms}\cdot \mathbf{a}^{-}({\rm R}), \nonumber
\end{eqnarray}

\noindent and from it we obtain the secular determinant:

\begin{eqnarray}
\label{r-dispersionS-N}
 \Det \left\{ \left[\bQ{\rm R}_{11}\cdot\bS_{21}-\bQ{\rm L}_{11}\;e^{iq d}-\bQ{\rm L}_{12}\cdot\bS_{11}\;e^{iq d}\right]^{-1}\right. \cdot \nonumber\\ \left.\left[\bQ{\rm L}_{12}\cdot \bS_{12}\;e^{iq d}-\bQ{\rm R}_{11}\cdot\bS_{22}-\bQ{\rm R}_{12}\right]\right.\nonumber\\\left. -\left[\bQ{\rm R}_{21}\cdot\bS_{21}-\bQ{\rm L}_{21}\;e^{iq d}-\bQ{\rm L}_{22}\cdot\bS_{11}\;e^{iq d}\right]^{-1}\right. \cdot \nonumber\\ \left.\left[\bQ{\rm L}_{22}\cdot\bS_{12}\;e^{iq d}-\bQ{\rm R}_{21}\cdot\bS_{22}-\bQ{\rm R}_{22}\right]\right\}&=& 0\;.
\end{eqnarray}

We note that the equations (\ref{r-dispersionH-N}),
(\ref{r-dispersionE-N}) and (\ref{r-dispersionS-N}) are given in
terms of matrix blocks that can overcome the numerical instability
known as $\Omega d$ problem. This is not the case for the secular
equation in terms of $\bT (z+d, z)$:

\begin{eqnarray}\label{r-dispersionT-N}
  \Det[\bT(z+d, z)-\bI\;e^{iqd}] &=& 0\;,
\end{eqnarray}

We shall consider now as an example the motion of electrons in a
periodic one-dimensional potential such as that of a superlattice
formed by barriers of $B$ material with effective mass $m_{B}$,
thickness $b$ and height $V_0$ and wells of $A$ material with
effective mass $m_{A}$ and thickness $a$. In this case the
equations (\ref{r-dispersionH-N}), (\ref{r-dispersionE-N}),
(\ref{r-dispersionS-N}) and (\ref{r-dispersionT-N}) are given by:

\begin{eqnarray}
 \label{H-uni}  2\cos(qd)\bH_{12} &=& 1-\bH_{11}\bH_{22}+ \bH_{12}^2 \\
 \label{E-uni} 2\cos (qd)\bE_{12} &=& \bE_{22}-\bE_{11} \\
 \label{S-uni} 2\cos(qd)\bS_{12} &=& \frac{k_{_{B}}m_{_{A}}}{k_{_{A}}m_{_{B}}}\left(\bS_{21}\bS_{12}-\bS_{11}\bS_{22}\right) +1 \\
  \label{T-uni} \cos (qd) &=& \frac{1}{2} (\bT_{11}+\bT_{22}),
\end{eqnarray}

where

\begin{eqnarray}
k_{A} & = & \sqrt{\frac{2m_{A}}{\hbar^{2}}E} \\ \\
k_{B} & = & \sqrt{\frac{2m_{B}}{\hbar^{2}}(E-V_{0}}).
\end{eqnarray}

We used the $\sin k (z-z_\ell/z_r)$ and $\cos k (z-z_\ell/z_r)$
base in the $\mathrm{L}/\mathrm{R}$ domain to obtain
(\ref{S-uni}). When we use the matrix elements of $\bT$ for this
problem in the period $d=a+b$ in (\ref{T-uni}) we arrive to the
well known Kronig-Penney equation
\cite{1931-KronigPenney-ProcRSocLond}. The expressions
(\ref{H-uni})-(\ref{S-uni}) are variations of this equation if we
notice that the matrices $\bH$, $\bE$ and $\bS$ can be calculated
from their relations with $\bT$.

Expressions (\ref{H-uni})-(\ref{S-uni}) are variations of
(\ref{T-uni}) to calculate the system energy levels for any
barrier width $b$. When the barrier thickness $b\rightarrow\infty$
(limit of isolated symmetric rectangular wells) the secular
equation in terms of $\bT$ diverges. On the other hand its
variations lead directly to the well known transcendental
equations giving the energy levels for  even and odd states of a
symmetric rectangular well of width $a$ and depth $V_0$.

After some algebra it can be shown that equation (5.37) coincides
with the equation (32) in \cite{Smulzw1} for the Kronig-Penney
equation. In the same way it coincides with the equation (20) of
Ref. \cite{Smulzw2}. Refs. \cite{Smulzw1,Smulzw2} give results for
the Kronig-Penney equation to avoid the $\Omega d$ problem.

\section{\label{Conclusiones}Conclusions}

In the general framework of $N$ equation systems of the
Sturm-Liouville matrix kind with piecewise constant coefficients
we have shown that there are transfer matrix variants with
dimensions independent of the number of layers in the structure
which can avoid the numerical instabilities present in the ATM.
The hybrid compliance-stiffness matrix and the scattering matrix
can avoid the so called $\Omega d$ problem, being numerically
stable independently of how big or small be the thicknesses in the
multilayer structure. The Stiffness matrix and its inverse the
compliance matrix are numerically stable for big thicknesses of
the layers or of the multilayer structure. On the other hand, in
the case of very small layer thicknesses these two matrices can
exhibit the $\Omega d$ problem due to the roundoff errors
accumulation.For zero thicknesses both matrices exhibit a
numerical singularity (overflow).

Given the big variety of boundary problems which can be studied
with these numerically stable variants of the ATM and the
generality and ubiquity of the matrix Sturm-Liouville system, the
results obtained here can be applied to the study of various
elementary excitations in multilayer systems.

The relations between the different matrices studied here has
proven to be an useful instrument in the study of the numerical
stability of transfer matrices. With this technique it was
possible to show analytically the capability of some of these
variants of the transfer matrix to avoid the numerical degradation
leading to the $\Omega d$ problem.

In recent years some methods able to deal with systems having
inhomogeneous coefficients have been developed. We present in
Appendix C the link of the $N$ equation systems of the
Sturm-Liouville matrix kind to the corresponding differential
forms of those problems.

\appendix

\section{\label{App-A} Example of a matrix subset with a similar behaviour from the numerical point of view}

Let us denote by $\bX$, $\bY$, $\bZ$ and $\bR$ four matrices in
whose definition enter the vectors $\mathbf{F}(z)$,
$\mathbf{F}(z_0)$, $\mathbf{A}(z)$ and $\mathbf{A}(z_0)$, as for
example:

\begin{eqnarray}
  \begin{array}{cc}
    \begin{array}{|c|}
     \mathbf{A}(z_0) \\
     \mathbf{F}(z)
   \end{array} = \bX\cdot\begin{array}{|c|}
                               \mathbf{F}(z_0) \\
                               \mathbf{A}(z)
                             \end{array} & \begin{array}{|c|}
                                                 \mathbf{F}(z) \\
                                                  \mathbf{A}(z_0)
                                                    \end{array} = \bY\cdot\begin{array}{|c|}
                                                                           \mathbf{F}(z_0) \\
                                                                           \mathbf{A}(z)
                                                                            \end{array} \\
   &\\
    \begin{array}{|c|}
     \mathbf{A}(z_0) \\
       \mathbf{F}(z)
   \end{array} = \bZ\cdot\begin{array}{|c|}
                               \mathbf{A}(z) \\
                                \mathbf{F}(z_0)
                             \end{array} & \begin{array}{|c|}
                                                   \mathbf{F}(z) \\
                                                   \mathbf{A}(z_0)
                                                   \end{array} = \bR\cdot\begin{array}{|c|}
                                                                       \mathbf{A}(z) \\
                                                                        \mathbf{F}(z_0)
                                                                         \end{array}
  \end{array}
\end{eqnarray}

If we take as the reference matrix any one of them it can be shown
that one of the remaining matrices is obtained by permutations
among them of the vectors in the right-hand side of the reference
matrix. A second one is obtained by following this method among
the vectors in the left-hand side of the reference matrix. Finally
the third one is obtained with both permutations. The relations
between these matrices can be resumed as:

\begin{eqnarray}
  (\bX)_{11} &=& (\bY)_{21}=(\bR)_{22}=(\bZ)_{12} \\
  (\bX)_{12} &=& (\bY)_{22}=(\bR)_{21}=(\bZ)_{11} \\
  (\bX)_{21} &=& (\bY)_{11}=(\bR)_{12}=(\bZ)_{22}\\
  (\bX)_{22} &=& (\bY)_{12}=(\bR)_{11}=(\bZ)_{21}
\end{eqnarray}

These relations show that the matrices $\bX$, $\bY$, $\bZ$ and
$\bR$ will have a similar behaviour from the numerical point of
view.

\section{\label{App-B} Matrix $\bT$ partitions of order $N$}

Starting with the expression:

\begin{eqnarray}
\label{T-d-Apend}
   \bT(d) &=& \left[
                   \begin{array}{cc}
                       \mathbf{F}_{0_N} &\;  \mathbf{F}_{0_{2N}}\\
                      \mathbf{A}_{0_N} &\;\mathbf{A}_{0_{2N}} \\
                   \end{array}
                 \right]\cdot \left[ \begin{array}{cc}
                                \mathbf{\Pi}_{k_N}(d) & \bcero \\
                                \bcero  & \mathbf{\Pi}_{k_{2N}}(d)
                              \end{array}\right] \cdot \left[
                   \begin{array}{cc}
                       \mathbf{F}_{0_N} &\;  \mathbf{F}_{0_{2N}}\\
                      \mathbf{A}_{0_N} &\;\mathbf{A}_{0_{2N}} \\
                   \end{array}
                 \right]^{-1},
\end{eqnarray}

\noindent we have:

\begin{eqnarray}\label{particiones-T}
   \bT_{11} &=& \mathbf{F}_{0_{N}}\cdot\mathbf{\Pi}_{k_N}(d)\cdot\gamma_{11}^{-1}+\mathbf{F}_{0_{2N}}\cdot\mathbf{\Pi}_{k_{2N}}(d)\gamma_{12}^{-1}.  \nonumber\\
  \bT_{12} &=& \mathbf{F}_{0_{N}}\cdot\mathbf{\Pi}_{k_N}(d)\cdot\gamma_{21}^{-1}+\mathbf{F}_{0_{2N}}\cdot\mathbf{\Pi}_{k_{2N}}(d)\gamma_{22}^{-1}.  \nonumber\\
   \bT_{21} &=& \mathbf{A}_{0_{N}}\cdot\mathbf{\Pi}_{k_N}(d)\cdot\gamma_{11}^{-1}+\mathbf{A}_{0_{2N}}\cdot\mathbf{\Pi}_{k_{2N}}(d)\gamma_{12}^{-1}.  \nonumber\\
   \bT_{22} &=& \mathbf{A}_{0_{N}}\cdot\mathbf{\Pi}_{k_N}(d)\cdot\gamma_{21}^{-1}+\mathbf{A}_{0_{2N}}\cdot\mathbf{\Pi}_{k_{2N}}(d)\gamma_{22}^{-1}.
\end{eqnarray}

\begin{eqnarray}
  \gamma_{11} &=& [\mathbf{F}_{0_{N}}-\mathbf{F}_{0_{2N}}\cdot\mathbf{A}_{0_{2N}}^{-1}\cdot \mathbf{A}_{0_{N}}]. \nonumber\\
  \gamma_{12} &=& [\mathbf{F}_{0_{2N}}-\mathbf{F}_{0_{N}}\cdot\mathbf{A}_{0_{N}}^{-1}\cdot\mathbf{A}_{0_{2N}}]. \nonumber\\
  \gamma_{21} &=& [\mathbf{A}_{0_{N}}-\mathbf{A}_{0_{2N}}\cdot\mathbf{F}_{0_{2N}}^{-1}\cdot\mathbf{F}_{0_{N}}]. \nonumber\\
  \gamma_{22} &=& [\mathbf{A}_{0_{2N}}-\mathbf{A}_{0_{N}}\cdot\mathbf{F}_{0_{N}}^{-1}\cdot\mathbf{F}_{0_{2N}}].
\end{eqnarray}

\section{\label{App-C} Sturm-Liouville matrix form for inhomogeneous media}

The matrix Sturm-Liouville equation

\begin{equation}\label{ESLM}
\frac{d}{dz} \left[ \bB(z) \cdot \frac{d \bF(z)}{dz} + \bP(z) \cdot \bF(z)\right] + \bY(z)
\cdot \frac{d\bF(z)}{dz} + \bW(z) \cdot \bF(z)  = {\mathbf{L}}(z)\cdot \bF(z)=\bcero_{N\times 1} \;,
\end{equation}

can be written as:
\begin{eqnarray}\label{Sextic-formalism}
\\
\nonumber
  \frac{d}{dz}\left|\begin{array}{c}
                  \bF(z) \\
                  \bA(z) \\
                \end{array}\right| &=& \left(\begin{array}{cc}
                                           -\bB(z)^{-1}\bP(z) &\;\; \bB(z)^{-1} \\
                                           \bY(z)\cdot\bB(z)^{-1}\cdot\bP(z)-\bW(z) &\;\; -\bY(z)\cdot\bB(z)^{-1}\\
                                         \end{array} \right)\cdot\left|\begin{array}{c}
                                                                              \bF(z) \\
                                                                              \bA(z) \\
                                                                              \end{array}\right|.
\end{eqnarray}

Here $\bA(z)=\bB(z) \cdot \displaystyle{\frac{d \bF(z)}{dz}} + \bP(z) \cdot \bF(z)$ is the SLM operator matrix differential form.

The equation (\ref{Sextic-formalism}) is the link with the first order differential equations systems given in
eq.(2.10) of \cite{2003-Shuvalov-PRSL}, eq.(3.3) of \cite{2010-Norris-QJMAP}, eq.(2) and (A.6) of \cite{2013-Norris-JSV},
eq.(3) of \cite{2013-Korotyaeva-APL} and eq.(8) of \cite{2014-Korotyaeva-JCA}.

These equations cover different inhomogeneous systems.

\subsection{Radially inhomogeneous cylindrically anisotropic systems}

This is the case considered in \cite{2003-Shuvalov-PRSL}. In this
work the mass density and the elements of the stiffness tensor
depend only on the radial coordinate $r$. It is then possible to
write:

\begin{equation}\label{}
    \bu=C\bU^{(n)}(r) e^{i(n\theta+\kappa_z-\omega t)},
\end{equation}
\noindent to obtain:
\begin{eqnarray} \label{ESLM-cilindrica-elastico}
  \frac{d}{dr}\left[r{\mathbf{\hat{Q}}} \frac{d(C\bU)}{dr}+(\bR\bk+i\kappa_z
 r {\mathbf{\hat{P}}})\right]+\left(\bk\bR^T+i\kappa_z r
{\mathbf{\hat{P}}}^T\right)\frac{d(C\bU)}{dr} \nonumber\\
+\frac{1}{r}\left[\bk {\mathbf{\hat{T}}}\bk+ i\kappa_z r (\bk
{\mathbf{\hat{S}}}+{\mathbf{\hat{S}}}^T \bk)+ (i\kappa_z
r)^2({\mathbf{\hat{M}}}-\bI\rho\omega^2/\kappa_z^2)\right](C\bU)
&=& 0.
\end{eqnarray}

Here $C$ is a normalization constant. Equation
(\ref{ESLM-cilindrica-elastico}) is of kind (\ref{ESLM}) for a
cylindrical elastic material radially inhomogeneous. Here $r$
plays the role of $z$ in the planar systems. With the properties
imposed on $\bk$, ${\mathbf{\hat{Q}}}$, ${\mathbf{\hat{P}}}$,
$\bR$, ${\mathbf{\hat{S}}}$, ${\mathbf{\hat{T}}}$ and
${\mathbf{\hat{M}}}$ in \cite{2003-Shuvalov-PRSL} we have
$\bY=-\bP^{\dag}$, $\bB=\bB^{\dag}$ and $\bW=\bW^{\dag}$.

In this case the linear differential form in
(\ref{ESLM-cilindrica-elastico}) is $\bA=r \mathbf{t}_r$ (where
$\mathbf{t}_r)$ is the stress radial component and from eq.(2.7)
from \cite{2003-Shuvalov-PRSL} we obtain
$\bA(r)=Cr\bUpsilon^{(n)}(r)$. We see also that $\bF(r) \equiv
C\bU^{(n)}(r)$. By identifying the $\bB$, $\bP$, $\bY$ and $\bW$
matrices in (\ref{ESLM-cilindrica-elastico}) and substitution in
(\ref{Sextic-formalism}) we arrive to eq.(2.10) of
\cite{2003-Shuvalov-PRSL}, eq.(3.3) of \cite{2010-Norris-QJMAP}
and eq.(A.6) of \cite{2013-Norris-JSV}:

\begin{eqnarray}
  \frac{d}{dr}\bEta(r)^{(n)} &=& \frac{i}{r} \bG(r)\bEta(r)^{(n)};
\end{eqnarray}

\begin{eqnarray}
  \bEta(r)^{(n)} &=& C \left( \begin{array}{c}
                          \bU^{(n)}(r) \\
                          i r \bUpsilon^{(n)}(r) \\
                        \end{array}\right);
\end{eqnarray}
\begin{eqnarray}
  \frac{i}{r} \bG(r) &=&  \left(\begin{array}{cc}
                                           -\bB(r)^{-1}\bP(r) &\;\; \bB(r)^{-1} \\
                                           \bY(r)\cdot\bB(r)^{-1}\cdot\bP(r)-\bW(r) &\;\; -\bY(r)\cdot\bB(r)^{-1}\\
                                         \end{array} \right).
\end{eqnarray}

\subsection{Shear-horizontal elastic waves in phononic crystals formed by inhomogenoeus anisotropic materials. Cartesian coordinates}

This is studied in \cite{2014-Korotyaeva-JCA} where the
displacement $\bu$ depends on $x_1$ and $x_2$, but after expanding
$x_1$ in plane waves they obtain the following ordinary
differential equation in $x_2$

\begin{eqnarray}
  -(\pmb{\partial}_{1}+i\kappa_1)(\bmu(\pmb{\partial}_{1}+i\kappa_1)\bu)+\partial_{2}(\bmu\partial_2\bu) &=& -\brho\omega^2\bu.
\end{eqnarray}

We can then identify $\bF(x_2)=\bu(x_2)$ and
$\bA(x_2)=\bmu\partial_2\bu$, $\bP=\bcero$, $\bY=\bcero$,
$\bB=\bmu^{-1}$ and
$\bW=\brho\omega^2-(\pmb{\partial}_{1}+i\kappa_1)
\bmu(\pmb{\partial}_{1}+i\kappa_1)$. After substitution of these
expressions in (\ref{Sextic-formalism}) $\bW$ acts on the
displacement $\bu(x_2)$ and we obtain eq.(8) of
\cite{2014-Korotyaeva-JCA} which is essentially the same than
eq.(3) of \cite{2013-Korotyaeva-APL}.

We have seen that in all these cases involving inhomogeneous
elastic anisotropic media we can put the matrix Sturm-Liouville in
the (\ref{Sextic-formalism}) form. Then it would be possible to
apply the stable integration methods of \cite{2003-Shuvalov-PRSL,
2010-Norris-QJMAP, 2013-Norris-JSV} for cylindrical geometry and
those of \cite{2013-Korotyaeva-APL, 2014-Korotyaeva-JCA} for
layered systems.

\thanks{\emph{Acknowledgments}. We thank the Associate Editor and
the referees for valuable comments and useful suggestions.}

\bibliographystyle{unsrt}
\bibliography{BibRPS-libros,BibRPS-articulos}

\end{document}